\RequirePackage{pdf14}% Also disables \pdfobjcompresslevel

\documentclass[10pt]{iopart}

\usepackage[pdftex]{graphicx} % For PDF figures
\graphicspath{{figures/}}
\DeclareGraphicsExtensions{.pdf,.jpeg,.png}

\usepackage[font={small}]{caption} % Make captions smaller
 
\usepackage{subfig} 
\usepackage{comment}
\usepackage{tabularx}
\usepackage[bottom]{footmisc} % Footnotes
\usepackage{makecell}
\usepackage[export]{adjustbox} % Use valign with includegraphics
\usepackage{array}
\newcolumntype{P}[1]{>{\raggedright\arraybackslash}p{#1}}
\newcolumntype{C}[1]{>{\centering\arraybackslash}p{#1}}
 
\usepackage{color, soul} % For highlighting of text 
 
\begin{document}

\title[]{The Engineering of a Scalable Multi-Site Communications System Utilizing Quantum Key Distribution (QKD)}

\author{Piotr~K.~Tysowski$^1$, Xinhua~Ling$^2$, Norbert~L{\"u}tkenhaus$^3$, and~Michele~Mosca$^4$}
\address{$^{1,3,4}$ Institute for Quantum Computing (IQC), University of Waterloo, 200~University Ave. W., Waterloo, Ontario, N2L 3G1, Canada}
\address{$^{2}$ XLNTec Inc., 329~Deerfoot Trail, Waterloo, Ontario, N2K 0B4, Canada}
\address{$^{4}$ Department of Combinatorics and Optimization, University of Waterloo, 200~University Ave. W., Waterloo, Ontario, N2L 3G1, Canada}
\address{$^{4}$ Perimeter Institute for Theoretical Physics, 31 Caroline St. N., Waterloo, Ontario, N2L 2Y5, Canada}
\address{$^{4}$ Canadian Institute for Advanced Research, 180 Dundas St. W., Toronto, Ontario, M5G 1Z8, Canada}
\address{$^{3,4}$ evolutionQ Inc., 295 Hagey Blvd., Waterloo, Ontario, N2L 6R5, Canada}
\ead{\mailto{$^1$p.tysowski@uwaterloo.ca}, \mailto{$^2$xinhua@xlntec.com},  \mailto{$^3$lutkenhaus.office@uwaterloo.ca},  \mailto{$^4$michele.mosca@uwaterloo.ca}}
\vspace{10pt}
%\begin{indented}
%\item[]February 2014
%\end{indented}

\begin{abstract}
\hbadness 10000 % Remove underfull \hbox warnings
Quantum Key Distribution (QKD) is a means of generating keys between a pair of computing hosts that is theoretically secure against cryptanalysis, even by a quantum computer. Although there is much active research into improving the QKD technology itself, there is still significant work to be done to apply engineering methodology and determine how it can be practically built to scale within an enterprise IT environment. Significant challenges exist in building a practical key management service for use in a metropolitan network. QKD is generally a point-to-point technique only and is subject to steep performance constraints. The integration of QKD into enterprise-level computing has been researched, to enable quantum-safe communication. A novel method for constructing a key management service is presented that allows arbitrary computing hosts on one site to establish multiple secure communication sessions with the hosts of another site. A key exchange protocol is proposed where symmetric private keys are granted to hosts while satisfying the scalability needs of an enterprise population of users. The key management service operates within a layered architectural style that is able to interoperate with various underlying QKD implementations. Variable levels of security for the host population are enforced through a policy engine. A network layer provides key generation across a network of nodes connected by quantum links. Scheduling and routing functionality allows quantum key material to be relayed across trusted nodes. Optimizations are performed to match the real-time host demand for key material with the capacity afforded by the infrastructure. The result is a flexible and scalable architecture that is suitable for enterprise use and independent of any specific QKD technology.
\end{abstract}


%\maketitle

\section{Introduction}
\label{sec:introduction}

\subsection{Background}

Secure network communication is a principal function of IT infrastructure. In particular, secure inter-domain communication is important to larger organizations that are distributed across multiple geographical sites. Computing hosts on a local site will regularly establish communication sessions with arbitrary computing hosts of a remote site. The key management service of an IT security system will typically provide a data encryption service on top of a standard network protocol. Although there are many commercially available crypto-systems in use today, the majority of them rely upon key exchange under public-key cryptography where the computational problem is infeasible to break using today's computing technology. However, rapid advances are occurring in the field of quantum computing. Once a quantum computer is built to solve problems of a practical scale, currently-deployed conventional public-key cryptography will become completely vulnerable to attack, and a new way of protecting transactions over a network will be needed. 

New quantum-resistant public-key cryptography, also known as post-quantum public-key cryptography, is being researched and experimented on, and there is growing effort toward its global standardization. While promising, and an important part of the future cryptographic landscape, post-quantum cryptography is still potentially vulnerable to future (quantum and classical) algorithmic advances. In particular, there has been limited scrutiny against novel quantum cryptanalysis, especially considering the lack of a large-scale quantum computer to facilitate the design and testing of new quantum algorithms and heuristics. Furthermore, post-quantum key establishment is susceptible to being recorded and cracked at a future date.

To mitigate the risk of successful attack, Quantum Key Distribution (QKD) has been devised, which is based on the laws of quantum physics and is a theoretically secure form of generating keys that is resistant to attacks by both conventional and quantum computers. There has been significant research activity over recent years relating to the theory and implementation of QKD techniques. QKD is already in operation today in large-scale experiments using commercially-available equipment, and promising research has been presented on how trusted nodes can be utilized to build a larger system. However, there is still work to be done to determine how QKD key management can be practically utilized within an enterprise environment to flexibly serve a large host population, from a practical engineering perspective. QKD must evolve from point-to-point links to network designs that can support high user populations over wide geographical coverage \cite{diamanti16}.

\subsection{Our Contributions and Impact}

The overall goal of our work has been to show the feasibility of integrating QKD technology with classical networks, to uncover unique design challenges and propose flexible solutions, and to provide a holistic view of a system as a viable target for migration. We viewed the problem primarily from an engineering lens. Our aim has been to apply best practices in contemporary security, network, and software engineering, including leading industry practices, in the application of QKD to practical real-world systems.

Our main contribution is the design of a scalable QKD-based system that enables secure multi-site communication, and is compatible with various QKD technologies. We provide a scalable service to support enterprise-level secure traffic between arbitrary hosts in a large metropolitan network comprising sites that may be indirectly connected. The architecture consists of a full protocol stack, including an enterprise-level key management layer that manages and issues keys to hosts from a key pool, and a quantum network layer that performs quantum key generation via trusted nodes. To maximize efficiency, the key generation system dynamically adapts to changes in demand and network infrastructure based on real-time monitoring and prediction from history. Hosts are issued session keys to securely communicate over a conventional network, while quantum key generation occurs over quantum links that may form an independent topology. Integrations with standards such as TLS, IPSec, and Kerberos have also been studied. The results of the research included design artifacts to lay the groundwork for the implementation of a research test bed or a pilot system for industry use.

Our work has numerous impacts: it informs QKD practitioners and equipment designers of the operating requirements of such an enterprise system; it informs architects of key design choices and trade-offs to make when incorporating QKD technology into a communications system; and, it demonstrates to software engineers how client applications can be built to make use of secure QKD key material in transparent fashion. These insights aim to tackle what sometimes appear to be insurmountable obstacles in standards acceptance and widespread use of QKD, despite its compelling intrinsic security benefits. Our research shows how the integration of QKD has implications on the entire systems design, and we suggest new avenues for study in the areas of security and network engineering to make QKD practical for widespread use. We concur with the ultimate goal cited in \cite{sasaki17}: to realize a quantum-safe infrastructure in which post-quantum cryptography, QKD, and physical-layer cryptography will be integrated.

This article presents a high-level architecture and design of a QKD-enabled communications system, to show how enabling quantum-safe communication security poses unique challenges to traditional network engineering; furthermore, it demonstrates how a feasible solution requires significant interplay between the various networking domains, from security protocols at the user level to network management at the infrastructure level. In the rest of this Section~\ref{sec:introduction}, related literature on practical deployment of QKD through protocol and network design is discussed, followed by a model which describes the typical enterprise network where quantum-safe communications security is desired. In Section~\ref{sec:system_design}, the overall design is presented, consisting of a layered protocol stack; the functionality and responsibility of the various layers that make up the protocol stack are described in turn from top-to-bottom. In Section~\ref{sec:procedures}, the main procedures that run in the system within the various layers of the stack and spanning them are described, including: the issuance of keys to hosts, demand management, and synchronization of the quantum key pool in the service layer, and the various key generation, routing, scheduling, and relaying functions of the control and data planes of the network layer. In Section~\ref{sec:trust_and_security}, the trust and security models are described, and as well as options for integration with post-quantum algorithms to attain further security robustness. Finally, in Section~\ref{sec:recommendations}, the key lessons learned throughout the design process are summarized, and recommendations given for further work in this area. 

\subsection{Related Work}

A significant focus of research in the application of QKD technology has been to find ways of integrating it into conventional crypto-systems so as to ease the migration. Current literature on the use of QKD-generated key material for securing communication sessions delves into some options for integration with existing popular security protocols. For instance, integration with IPSec and TLS is accomplished in {\cite{mink09}} by replacing the calculated Diffie-Hellman shared secret or the master secret, respectively, with QKD-generated key material contained within a key database. The protocols periodically request fresh keying material and fall back to conventional key exchange if QKD is not operational. Another option for TLS integration is suggested in {\cite{elboukhari10}}, where the handshake protocol is modified so that QKD is carried out in the middle of the handshake, and a configuration protocol is added to agree on parameters such as the key length. Our own proposal relies upon a pre-established QKD channel such that QKD-generated key material is continuously made available for each party, and takes advantage of the existing pre-shared secret key standards to avoid significant re-work in the protocols; the transmitted fields relating to the pre-shared secret are appropriately modified so that both parties obtain the same quantum-generated key, and significant re-work to the protocols are avoided. Proposals in the current literature result in different trade-offs in efficiency and implementation complexity but they address the most basic scenario of point-to-point communication only; the works do not further show how to engineer such systems for practical use within an enterprise environment to serve a large and dynamic host population, and to flexibly support different key generation technologies; these themes are a key focus of our work. In {\cite{assche06}}, the implementation of a QKD crypto-system is proposed through a secret-key distillation protocol, which occurs in the QKD process itself. A pair of users negotiates various quantum transmission parameters and performs on-demand key generation; it is also noted that synchronization of the pools containing quantum key material may be lost. The protocol, however, is again not extrapolated from a pair of users to a network, and the synchronization logic between parties that seems necessary is not elaborated upon.

Some work has been done to study the architecture required for securing communications within a network scenario. In {\cite{pattaranantakul12}} and {\cite{pattaranantakul15}}, a conceptual framework is presented consisting of QKD, key management, and application layers. The key management layer implements the standard Key Management Interoperability Protocol (KMIP) {\cite{kmip09}}. The system relies upon a global key management service that provides service to hosts. A key caching protocol is proposed for transferring QKD-generated keys from the QKD layer to the key management layer. Multiple key management servers can transfer quantum-generated keys between themselves via Virtual Private Network (VPN) tunnels. The QKD-generated keys may be  transmitted from the source to the destination in a hop-by-hop manner along a path of fully-trusted peer nodes using an encrypted transfer protocol; each pair of nodes along the path encrypts the QKD key using existing pairwise symmetric keys. Finally, a routing protocol is used to find the best next hop for each node. Although the framework is useful, it lacks strategy for coping with resource constraints in the network and dealing with the resulting limitations on the key generation rate. A very similar high-level framework for e-governance applications is presented in {\cite{murali15}}, where the QKD layer relies upon a cloud-hosted quantum processor {\cite{bristol17}}. In contrast, our proposed key management service is fully decentralized to minimize bottlenecks; we provide a comprehensive design for issuing keys in a scalable and technology-agnostic manner; we provide concrete implementation strategies for making best use of the available network capacity, applying flexible security policies to serve the encryption needs of a large user base, and for synchronizing the key generation and issuance processes across multiple sites. In short, we have designed a QKD-based communications system suitable for enterprise use.

Other work has explored the impact of a larger QKD system on network engineering. A prototype European QKD network called SeCoQC (Secure Com\-munication based on Quantum Cryptography) has been proposed in \cite{peev09}, where backbone nodes form a network for key distribution and provide best-effort and guaranteed-rate services. A backbone node contains modules for key forwarding and routing, and a load-balancing policy is enforced based on a shortest-path calculation. A multi-layer protocol stack is also defined. We propose more extensive routing and forwarding functions that optimize key generation in highly-dynamic environments with constant host demand and network changes. Furthermore, we provide a mechanism for providing different levels of security service based on the available network capacity at any time. Finally, our session keys utilizing quantum key material are negotiated directly between hosts to ensure scalability. In \cite{tajima17}, a QKD platform was proposed with an architecture consisting of application, key supply, key management, and quantum layers. The platform was demonstrated on the Tokyo QKD network in the context of a smartphone system. A hybrid approach was proposed in which streaming data is encrypted with AES while the cryptographic key is periodically refreshed from the quantum key supply. Key relay is used for key generation across longer distances in the network. However, scalability aspects and the flexibility of having different operating modes in the system are not specifically addressed, while they are central to our work. In \cite{hughes13}, a trusted authority becomes a central node for multiplexed quantum communication; it provides hierarchical trust to clients in a hub-and-spoke topology. This arrangement is argued to be more scalable than general trusted QKD networks that rely upon a mesh of point-to-point links between all communicating entities; however, each client still requires its own link to the trusted authority. Our QKD network nodes are co-located with key management servers that provide quantum key material to local enterprise hosts on conventional local networks and thus require less dedicated infrastructure overall.

\subsection{The Network Model}

\begin{figure}[htbp]
\centering
\includegraphics[scale=0.60]{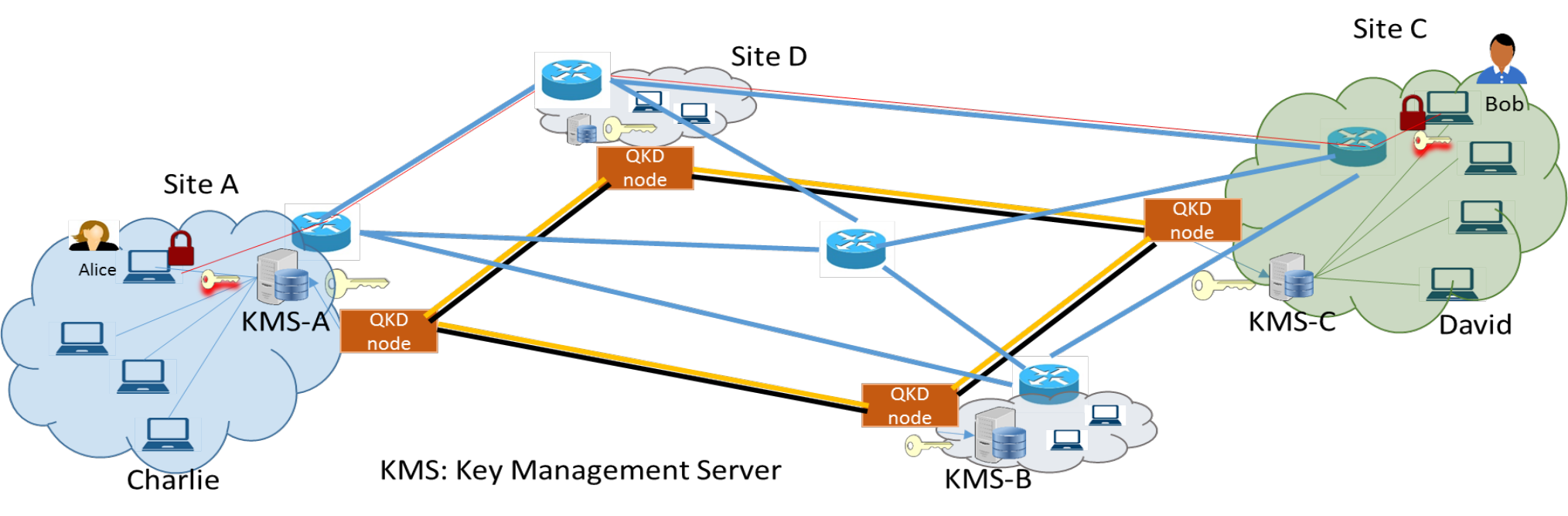}
\caption{The network model consists of multiple communicating sites, each containing hosts that request session keys from their local KMS (Key Management Service). Each KMS maintains a quantum key pool that contains QKD-generated key material. QKD is executed over a quantum network of such sites. Once hosts obtain session keys to encrypt individual communication sessions, they can engage in secure communication across sites over a conventional network connection, which may exhibit a different topology.}
\label{fig:network_model}
\end{figure}

\begin{figure}[htbp]
\centering
\subfloat[Pairwise communicating sites.]{{\includegraphics[scale=0.35]{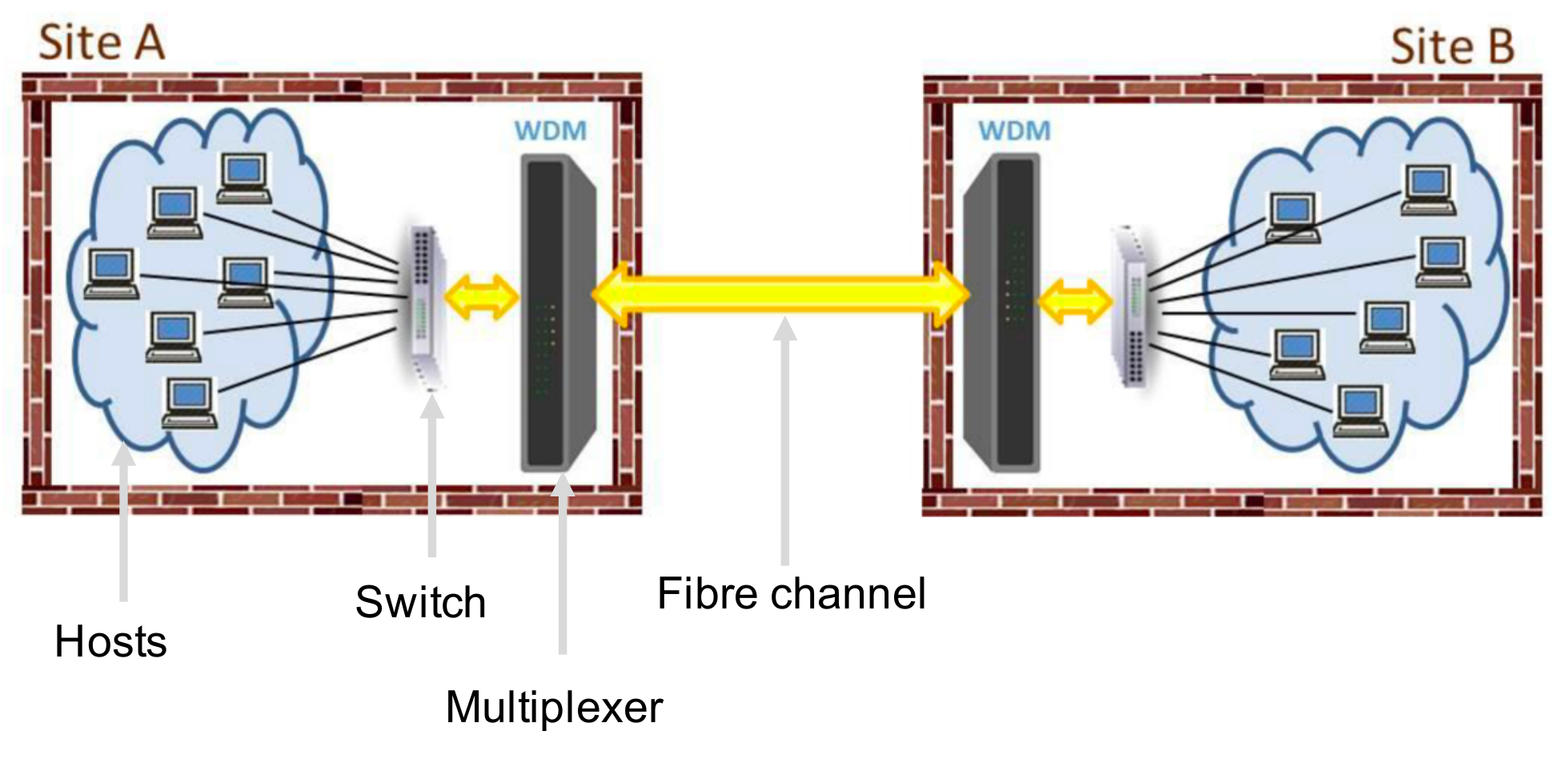}}}%
\quad
\subfloat[Multiple communicating sites.]{{\includegraphics[scale=0.35]{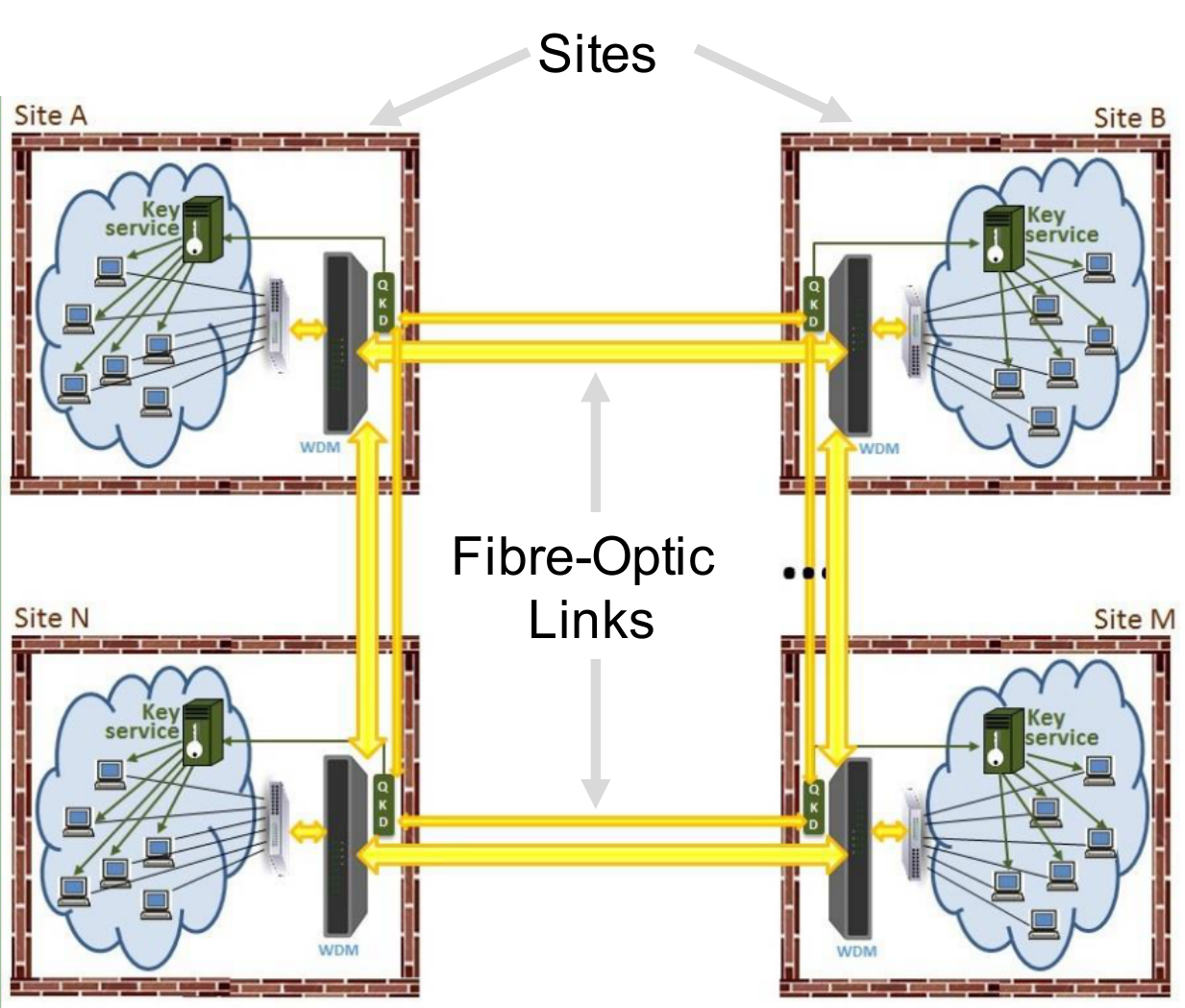}}}%
\caption{Pairwise sites typically communicate over an existing fibre-optic channel that is accessed by respective hosts through a switch and optical multiplexer employing wavelength-division multiplexing. Each site may be connected to more than one other site via an isolated optical path over which QKD is being executed on a dedicated pair of QKD devices; the collection of sites forms a quantum network.}
\label{fig:site_model}
\end{figure}

There is generally a need for a quantum-safe multi-site secure communications network for commercial entities as well as governments. The network model being addressed, as shown in Figure~\ref{fig:network_model}, is assumed to comprise multiple \emph{sites} that are geographically separated, such as buildings in a metropolitan area. Each site is a physically secure domain that contains a Local Area Network (LAN) consisting of potentially thousands of heterogeneous \emph{hosts}, or computing devices attached to the network, including desktop and mobile devices. Sites may be connected by fibre-optic channels with user data multiplexed through Wavelength-Division Multiplexing (WDM), and these are considered conventional traffic channels; the fibre is typically shared as opposed to being dark. LANs are connected to switches, and in turn, to Wavelength Division Multiplexers (WDMs), as shown in Figure~\ref{fig:site_model}. The fibre channels can be exploited to carry both secure user traffic as well as to perform QKD; thus, they form a   \emph{quantum network} consisting of \emph{quantum links} between each pair of communicating sites. The links between sites may not necessarily consist of fibre, however. The sites may alternatively be connected by any other channel enabling both quantum and classical communication, notably such as a free-space optical channel; the proposed architecture supports any such link. For the purpose of this discussion, it will be assumed that fibre channels are being utilized.

A fundamental difficulty is that the QKD protocol is designed to work for two parties only; however, in a metropolitan network, quantum-safe communication must be permitted between any arbitrary hosts on any arbitrary sites; a flexible addressing mechanism and scalability are therefore required. Additional challenges are the distances involved; sites may be separated by tens of kilometres, limiting the key generation rate. A metropolitan network may consist of tens of sites, and each site may contain thousands of hosts, so that there is great contention for quantum key material. Additionally, sites may not be fully mesh-connected by quantum links, so that relaying of quantum key material to remote sites is required. Finally, dedicated quantum links through dark fibre for QKD may not be available or cost-effective, so that existing fibre-optic lines for client traffic may need to be utilized for QKD. Each node\footnote{The terms \textit{site} and \textit{node} may be used almost interchangeably in the discussion of the metropolitan network; a site describes a physically secure domain housing a population of hosts utilizing secure communication services, and this facility becomes addressable as a node in inter-site data exchange over a conventional or quantum network.} in the network may have dual roles; it may act as an end-point for a connection or as a relay for the QKD key generation process, or both.

The host of each site retrieves a \emph{session key} from the locally-hosted Key Management Service (KMS); this key encrypts a single communication session between any two hosts located on separate sites. The same pair of users can engage in multiple secure sessions such as chat, file transfer, and videoconferencing, each requiring its own session key; this approach minimizes the impact of a single key compromise, and permits different algorithms and key types to be used for different applications as appropriate. Each KMS maintains a key repository, called a \emph{quantum key pool}, that contains key material generated through Quantum Key Distribution (QKD), referred to here as QKD-generated key material or \emph{quantum key material}. Once hosts obtain session keys to encrypt individual communication sessions, they can engage in secure communication across sites over a \emph{conventional network} connection, including a wireless broadband network, copper cable network, or long-haul optical fibre. The conventional network may have a different topology than that of the quantum network, and may include the use of routers and switches.

Other network engineering design options may be relevant to the QKD network. It is possible to have some QKD network backbone nodes, which are not associated with any specific sites, acting as relays only; these can improve network operation by adding possible paths and capacities. Note that conventional network devices cannot extend the range of the quantum channels used in QKD; transparent all-optical switching results in attenuation \cite{chapuran09}, while conventional routers and amplifiers are incompatible with QKD. Nevertheless, conventional passive optical switches can result in the creation of large and flexible network topologies for the purpose of executing QKD; this avoids sole reliance on a mesh of point-to-point links only where all nodes must function as relays for others. The routing logic in our design is compatible with any general approach to the topology. The scalability of the communications system is an important consideration that must be factored into applicable areas of its design, including processing, network, and storage capacities. As a very general example, a large enterprise network may reasonably be expected to consist of 5 to 20 sites, within a single major city, housing a total population of 10,000 to 20,000 hosts, with a total of 1,000 communication requests being generated per second. 

\section{The Design of a Scalable and Secure Communications System}
\label{sec:system_design}

\subsection{Layered Architectural Style}

Overall, we have designed a system for enterprise-level sites to securely communicate in a large metropolitan network. Key generation occurs using QKD technology over quantum channels connecting pairs of sites with maximum utilization of available capacity. A scalable service issues session keys from a quantum key pool, containing QKD-generated quantum key material, to local hosts on each site. Hosts can then use the session keys to securely communicate over conventional TCP/IP channels. The key generation and distribution mechanisms are designed to scale to many sites and hosts.

\begin{figure}[htbp]
\centering
\includegraphics[scale=0.50]{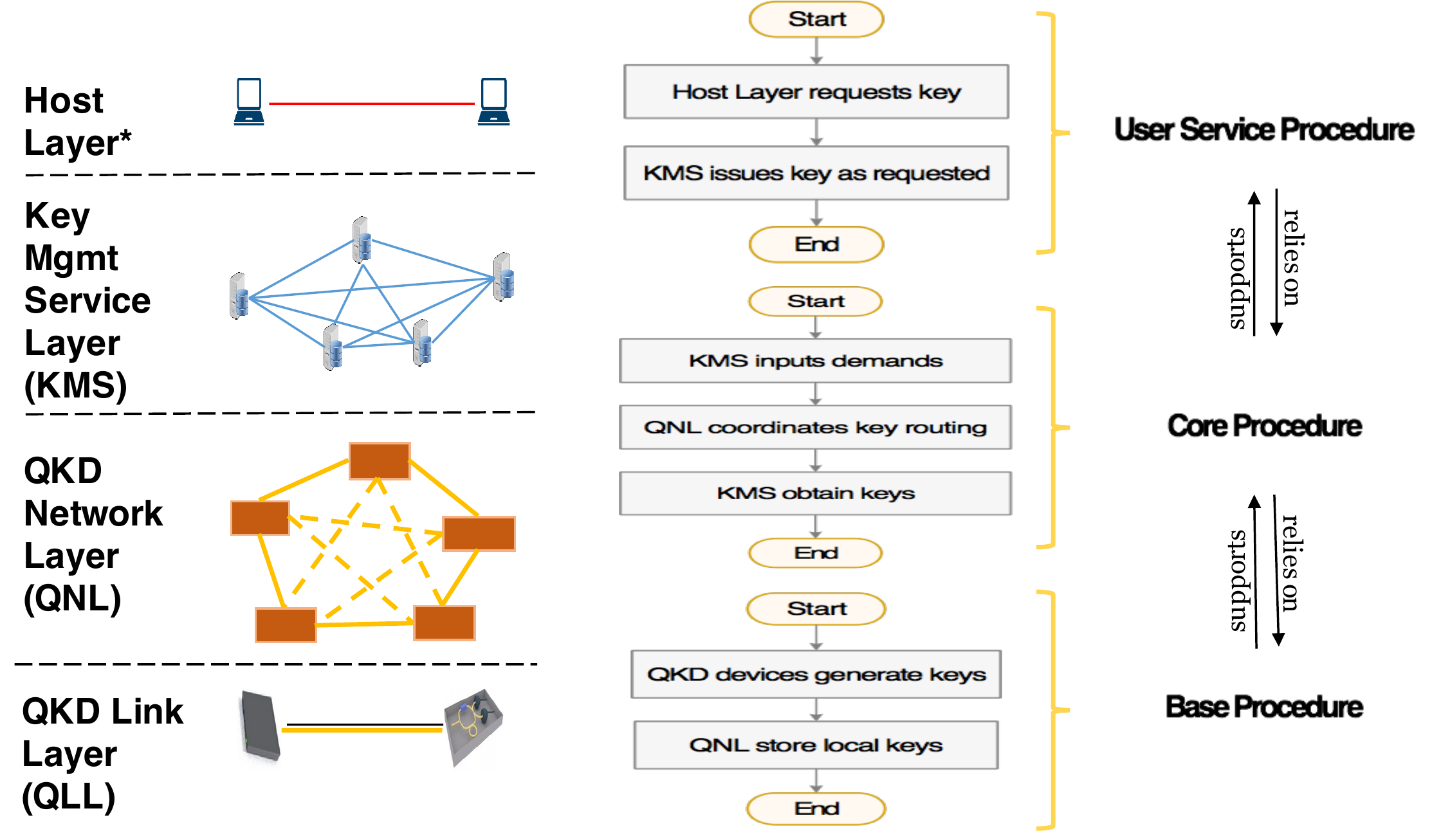}
\caption{The QKD communications system is composed of multiple layers. Various high-level procedures are constantly running across the stack. In the User Service Procedure, a host inside the Host Layer requests and is issued keys from the key pool within the Key Management Service (KMS) Layer. In the Core Procedure, the KMS layer in turn reports the demand for keys to the QKD Network Layer (QNL), which finds capacity in the quantum network to perform the key generation, and determines how the key generation data flows are to be routed through the trusted nodes. In the Base Procedure, the QNL then issues instructions to the Quantum Link Layer (QLL) to generate raw key bits along each route by employing QKD devices; these key bits are then passed up to the QNL, assembled as a key bit stream shared by the endpoints, and then passed up to the key pool in the KMS. There is a clear dependency between each pair of layers that is implemented by an interface contract.}
\label{fig:system_layers}
\end{figure}

The system is composed in a layered architecture style, as shown in Figure~\ref{fig:system_layers}; this software engineering model ensures that the major functions are grouped separately with well-defined interfaces across the layers that can be standardized. Traversing the protocol stack in the bottom-up direction, the Quantum Link Layer (QLL) produces raw key material by employing QKD hardware across a quantum link between two sites, based on instructions received from the Quantum Network Layer (QNL) above. The QNL issues key generation and routing requests for raw key material to be produced through the trusted-node network, and assembles and provides it to the Key Management Service (KMS) Layer above to fill its key pool. The KMS responds to key requests from the Host Layer above by issuing keys from its pool of QKD-generated key material that it manages, in accordance with a configured security policy. The Host Layer at the top encapsulates the software applications running on host hardware within a site, such as desktops and mobile devices, that require secure communication with other hosts; these applications request session keys, either through a proprietary or standard protocol.

The system optimizes key generation through a network of sites in which the majority would be expected to function as trusted nodes; these are sites which volunteer to act as intermediaries for other sites that do not share a direct quantum link over which key generation through QKD could occur. A trusted node engages in QKD with one site that is an endpoint in secure communication, and then in turn, the other endpoint; the trusted node then provides a translation between the two keys that it participates in generating, so that the endpoints are able to construct and share a single key in the end. A chain of trusted nodes can be built to cover larger geographical distances. Because each trusted node unavoidably knows the keys that it creates, it must be fully trusted by the endpoints. Although trusted nodes are the current state-of-the-art, there is very active research in developing practical quantum repeaters utilizing entanglement that will replace the need for trusted nodes, and will result in fewer compromises in the trust model.

The layered architecture results in a technology-independent design that can accommodate any QKD technology with minimal changes. Replacing fibre-link QKD with free-space QKD, for example, can occur just by modifying or replacing the lowermost link layer that communicates directly with the QKD devices. One can even design the option to use QKD in host applications irrespective of the current availability of QKD; should the service become available in the future, then those applications will be immediately enabled for it. This design choice can be made at the same time as post-quantum alternatives are designed into those applications. An implementation-level security policy can dictate which available technology to use to satisfy business needs. Application hosts can be made to choose an appropriate key agreement protocol such as SSL or IPSec based on engineering considerations; issues such as compatibility, ease of configuration, support for data compression, interoperability, and transfer speed are discussed in \cite{alshamsi05}; irrespective of the protocol chosen, QKD will generate quantum key material at a lower layer for its use. Further discussion on how QKD can be best combined with the use of post-quantum algorithms is found later in section~\ref{sec:combination_post_quantum}.

The following subsections examine the high-level responsibilities and functionalities of each individual layer in greater detail. The layers are presented in top-to-bottom order, to match the general direction of the control flow within the system. Then, in the next section, the operations of the system that enable interaction across the entire suite of layers are detailed.

\subsection{The Host Layer}
\label{sec:host_layer}

The Host Layer encompasses the end user applications engaging in secure information exchange required by various government and industry verticals. The applications run on a heterogenous mix of hardware including desktop computers, notebooks, and mobile devices. In particular, this layer may be manifested in software as a protocol stack in itself that will in its entirety act as a consumer of keys generated by the proposed QKD system; in this respect, the proposed system is agnostic to the security protocol used within the Host Layer. For example, one possible operating context is a mobile application running on a smartphone that utilizes Transport Layer Security (TLS) to open an encrypted connection to a service running on a web server situated on another site. As will be described later, TLS provides a mechanism that can be adapted to request keys from the proposed QKD system. TLS itself, or at least significant components of it, may be considered to map to the presentation layer of the OSI model (layer 6) or the application layer of the TCP/IP model; these mappings may be open to debate, but it is of no consequence here; for the purpose of describing the interaction of the user with the proposed QKD communications system, the mobile application and its use of TLS is considered to strictly occupy the Host Layer being defined here. The application will request QKD-generated keys from the Key Management Service layer below. Local access by hosts to this layer through the LAN is not restricted in the proposed solution; if warranted by the IT policy of the organization, then access controls such as Role-Based Access Control (RBAC) or one of many other controls may be additionally deployed, and may be integrated with the security policy enforcement function in this layer. Such local access controls would rely upon conventional, not quantum-safe, cryptographic techniques; they would strictly provide authentication internally within a site only.

\subsection{Optional Third-Party Local Key Management Service Layer}
\label{sec:local_kms_layer}

Optionally, a KMS (Key Management Service) layer, implemented by a third-party, may reside between the Host Layer and the KMS Layer of the proposed QKD communications system. Its existence depends on the nature of the deployment of the proposed QKD communications system, and is not depicted in Figure~\ref{fig:system_layers}; it may be part of a standardized or a custom solution. This layer is responsible for disseminating QKD-generated session keys obtained from the proposed QKD system via some additional mechanism to local hosts on one or more sites. For instance, the Kerberos network authentication system (the integration of which will be covered later in Section~\ref{subsec:standard_protocol_integration}) provides its own Key Distribution Centre (KDC) that works in concert with a Ticket-Granting Service (TGS) to issue session keys to clients. The TGS will obtain its session keys from the QKD system's KMS Layer below, rather than generating them itself; in this way, the QKD system's functionality is abstracted from the third-party key management service. Furthermore, the third-party KMS may fall back to using conventional cryptography if the QKD communications system is unavailable.

\subsection{The (QKD) Key Management Service Layer}
\label{sec:kms_layer}

\begin{figure}[htbp]
\centering
\subfloat[Key Management Service Layer]{{\includegraphics[width=0.45\textwidth,valign=bottom]{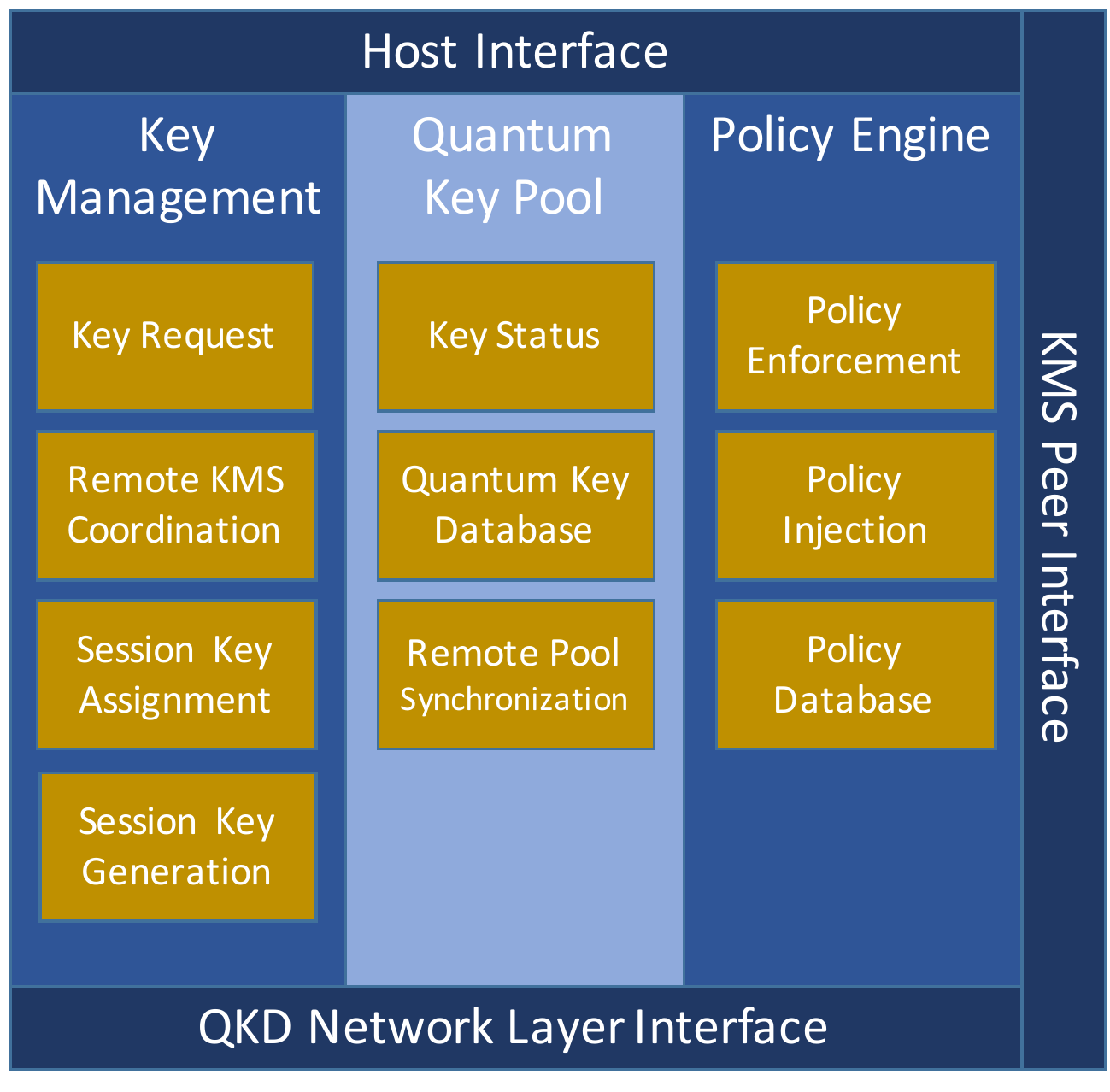}}\label{fig:kms_functional_view}}%
\hspace*{0.2cm}
\subfloat[QKD Network Layer]{{\includegraphics[width=0.50\textwidth,valign=bottom]{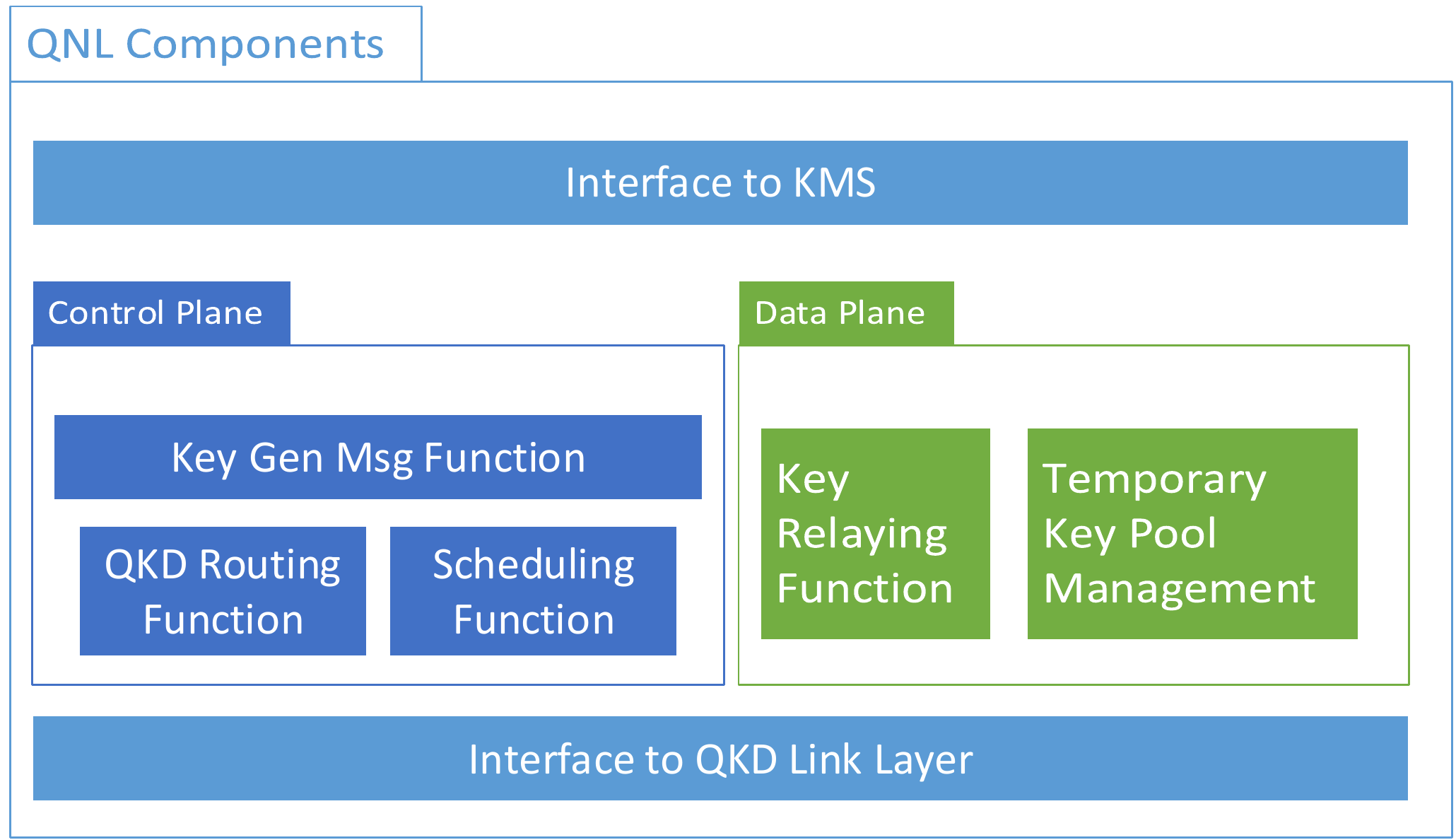}}\label{fig:qnl_functional_view}}
\caption{High-level functional views of the Key Management Service (KMS) Layer and QKD Network Layer (QNL) are shown. The KMS has three main areas of responsibility: key management, in which it constructs and issues session keys to local hosts from QKD-generated key material; maintenance of the Quantum Key Pool, which stores the key material that is kept synchronized with other sites, and a policy engine which enforces security policies on key issuance. The KMS interfaces with the immediately adjacent Host Layer and QNL, and also initializes QKD key generation with the other sites through a peer interface and a secured conventional network channel. The QNL has two main areas of responsibility: a control plane that schedules and manages key generation requests to other sites by finding the optimal routes for key generation through the trusted node network; a data plane that handles the flows of raw key data across trusted nodes, and allocates temporary key pools before final assembly of key material for an arbitrary pair of endpoints. The QNL interfaces with its own adjacent layers: the KMS above, through which it listens for key generation demands and then passes up its finalized raw key material for storage in the key pool; the QKD Link Layer below, through which it sends lower-level instructions so that QKD devices start or stop operation.}
\end{figure}

The Service Layer contains a KMS (Key Management Service) that issues QKD-generated keys to hosts to secure new communication sessions with hosts on other sites. The KMS reads quantum-generated key material from the network layer below, and manages the key material in a quantum key pool. At all times, the KMS maintains synchronization of the key pool with other sites. It constructs session keys for hosts using an appropriate key construction strategy that is consistent with the security policy contained in its policy engine; for instance, the policy may dictate different key lengths and lifetimes, depending on the cryptographic algorithm to be utilized by the host; suggestions for quantum-safe algorithms and suitable parameters are found in \cite{etsi2015}. The KMS will issue session keys to local hosts upon request while the quantum key pool contains sufficient key material. A limited key lifetime will require refresh of the key through a re-issue process similar to the original issue. As dictated by the policy, the KMS will make an appropriate response when the key pool is nearly exhausted; for example, it may wait for additional quantum key material to be fed by the network layer, or re-use existing material using a key expansion technique, such as the Rijndael key schedule \cite{fips197}. The KMS consists of several functional areas that may be decomposed into a modular software structure, as shown in Figure~\ref{fig:kms_functional_view} and described in detail in Table~\ref{tab:kms_functions_described}.

\begin{table}
\centering
\small
\begin{tabularx}{\textwidth}{|P{2.5 cm}|>{\raggedright\arraybackslash}X|} 
\hline \multicolumn{2}{|P{12.5 cm}|}{\bf  The Key Management Function within the KMS in the Service Layer is responsible for key management, and consists of the following modules:} \\
\hline Key Request module & As the primary interface point for hosts, this module services requests for session keys. It also collects and analyzes demand statistics. When the quantum key pool is exhausted, it falls back to key derivation if allowed by the policy, or blocks the host. \\
\hline Remote KMS Coordination module & This module coordinates with the remote KMS over the conventional network to start the process of key generation through the network layer. It also terminates the process for maintenance or error recovery.\\
\hline Session Key Assignment module & This module assigns session keys from the quantum key pool if sufficient bits are available. \\
\hline Session Key Generation module & This module starts and stops the key generation process by notifying the network layer of its local host demand, and will suggest a key generation rate; it will also initialize the quantum key pool size based on the projected peak demand. \\
\hline \multicolumn{2}{|P{12.5 cm}|}{\bf  The Quantum Key Pool Function is responsible for the management of the quantum key pool, and consists of the following modules:} \\
\hline Key Status module & This module keeps track of the current status of key bits as keys are assigned to fulfill session key requests. Key bits will need to be reserved (even after allocation to the local host) until the remote host also retrieves the same key, and the key negotiation protocol concludes. \\
\hline Quantum Key Database module & This module contains a copy of available QKD-generated key bits, also referred to as quantum key bits, read from the QKD network layer. The key bits, once populated in this database, are guaranteed to be synchronized with the remote site. \\
\hline Remote Pool Synchronization module & This module ensures that the key pool database on the remote site is synchronized at all times with the local database. This module will handle various exceptional conditions which may require purging the database and halting the QKD process. \\
\hline \multicolumn{2}{|P{12.5 cm}|}{\bf  The Policy Engine Function is responsible for enforcing policy rules governing key generation and use, and consists of the following modules:} \\
\hline Policy Enforcement module & This module verifies the security policy requirements on the hosts and keys. Rules are enforced during key assignment, such as those pertaining to key length and lifetime. \\
\hline Policy Injection module & This module allows an administrator to inject a policy into the policy database. \\ 
\hline Policy Database module & This module contains all of the security policies. Different policies may be defined for different classes of users and resources. \\
\hline \multicolumn{2}{|P{12.5 cm}|}{\bf  The external interfaces to the KMS, and the KMS modules that communicate via these interfaces, consist of the following:} \\
\hline Host interface & This interface accepts key requests from hosts in the Host Layer by the Key Request module. The interface may be implemented using an appropriate remote method invocation that is externally accessible. \\
\hline QKD Network Layer interface & This interface reads quantum key bits from the QKD Network Layer and stores them in the Quantum Key Database. The network layer is typically accessed through an internal call into the immediate layer below the current Service Layer within the protocol stack. \\
\hline KMS Peer interface & This interface is utilized by the Remote KMS Coordination module for synchronization with the KMS on the remote site, and may be implemented by a TCP/IP socket over a network. \\
\hline 
\end{tabularx}
\caption{Descriptions of the high-level functions of the Service Layer.}
\label{tab:kms_functions_described}
\end{table}

\begin{figure}[htbp]
\begin{center}
\includegraphics[scale=0.40]{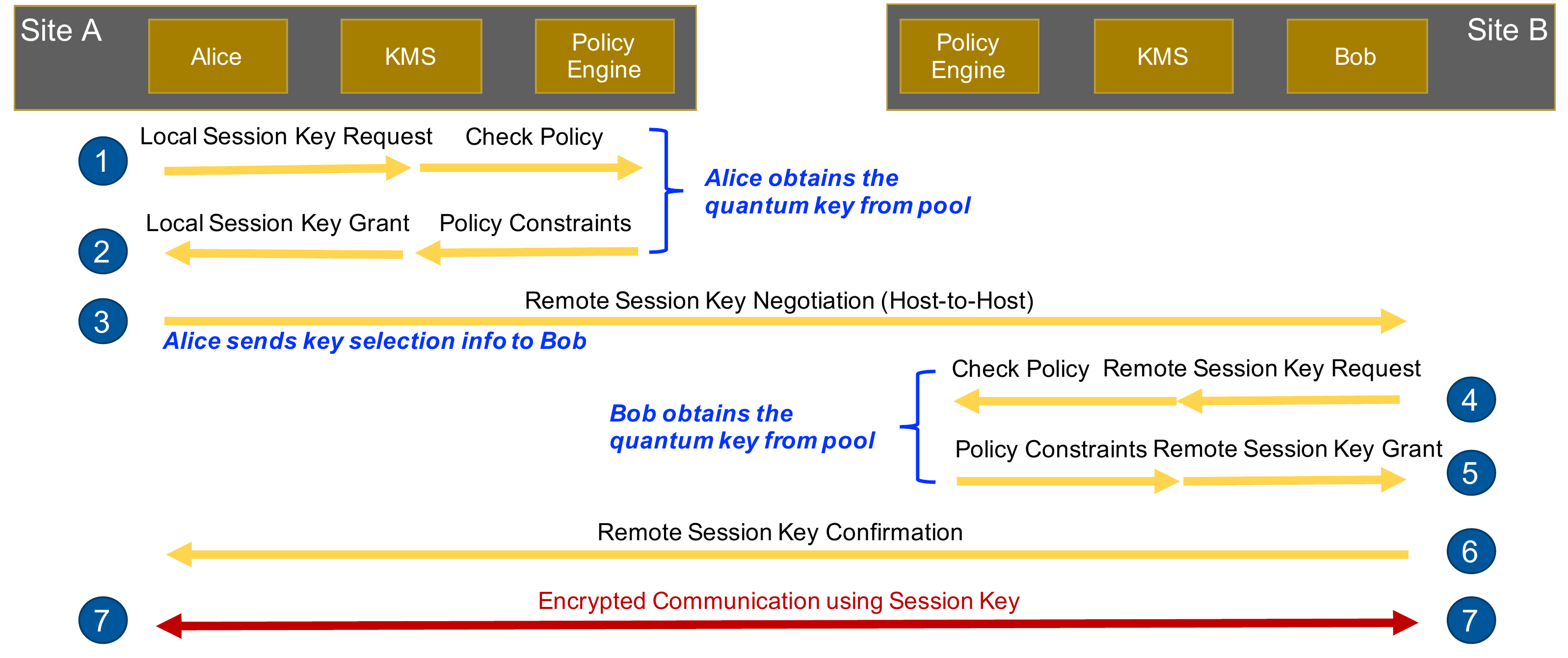}
\end{center}
\centering
\caption{The generic protocol for key negotiation between Alice on Site~A and Bob on Site~B is shown with a high-level representation of the actual message exchange. Alice requests a session key from the local KMS in order to begin secure communication with Bob. Alice's local KMS enforces any relevant policies, and grants a session key to Alice with an optional expiry time, as well as a signed packet containing key selection information. Alice then contacts Bob directly over the conventional network and provides this packet so that Bob can refer his own KMS to retrieve the same key; this is possible because the key pools of both sites are synchronized with the same content. Alice and Bob can then begin secure communication with the same symmetric session key. This key will be periodically refreshed by the hosts if necessary through an exchange with the KMS similar to the key grant scenario; the refresh requirements are specified by the KMS during key grant.}
\label{fig:key_negotiation}
\end{figure}

The KMS can issue keys to hosts using a proposed generic protocol as shown at high-level in Figure~\ref{fig:key_negotiation}. To ensure scalability, minimal server state is maintained by the KMS. The host is responsible for negotiating a secure session after retrieval of the session key. The crucial characteristic is that for hosts Alice and Bob communicating from separate sites, Alice's KMS issues a session key to her; Alice then transmits key selection information to Bob so that he may retrieve the same session key from his own KMS. Because the quantum key pools are always in sync during the QKD process, and the selection information is an index into the pool, the session key itself is never actually transmitted and thus cannot be eavesdropped. Within a single site, the KMS can be hosted as a virtualized service on a rack or blade server to achieve scalability. The server should be physically secured so that there is no direct access to the quantum key pool.

It is also possible to utilize a variant of this generic protocol where the local and remote KMSs, rather than the hosts, directly interact during key agreement. After the initial key grant to the originating host, Alice's KMS will directly transmit the key selection information to Bob's KMS through the authenticated KMS peer-to-peer interface. Although this results in slightly greater complexity in the KMS implementation and requires limited state information on the session to be kept server-side, it reduces host-to-host communication which can be costly especially in a wireless context; furthermore, it allows both KMSs to perform host authentication, and the remote KMS can notify the remote host of the incoming session through an efficient push-based notification mechanism. The process of agreement of QKD-generated keys has also been integrated with standard security protocols such as TLS; the integration approach is discussed later in Section~\ref{subsec:standard_protocol_integration}.

The KMS ensures continuous synchronization of its quantum key pool with that of the corresponding remote site for each site-to-site connection. The key pools must always be in sync so that the same quantum key material can be referenced and discovered during session key negotiation conducted by the sites' hosts. The KMS's coordinate securely through a communications channel that is typically a conventional network; the connection is authenticated through a mechanism such as that which will be described later in Section~\ref{sec:authentication_sites}. Another possibility is to have each KMS make a request into its respective underlying network layer, and all KMS coordination occurs through that layer.

The KMS Layer communicates with the Network Layer below to extract the quantum key material for storage in the quantum key pool. The KMS also initializes and terminates the overall site-to-site connection process. The underlying network layer is normally assumed to perform QKD key generation, but if the performance of this process is deemed insufficient or is interrupted, then another crypto-system can be utilized as a fall-back mechanism if allowed by the security policy. The other crypto-system may be based on another quantum-safe, or even a classical, algorithm and protocol, while its implementation maintains the same interface to the KMS layer.

\subsection{The QKD Network Layer}

The QKD Network Layer (QNL) provides quantum key material to the local KMS and negotiates key generation with other sites to enable inter-site communication. The QNL can flexibly contend with any network topology; for instance, some end nodes may not be directly connected as neighbours through a point-to-point quantum link; such nodes could be connected via more than one hop in a given network topology such as ring, star, or a partially meshed topology. In this case, the QNL needs to relay the key material via trusted nodes; in so doing, the QNL effectively forms a quantum network so that it can generate key material for any arbitrary pair of nodes. 

The QNL performs network control and management of the nodes via a \emph{control plane} technology suite, while key generation is undertaken by a \emph{data plane} technology suite. The data plane activities are controlled by the control plane technologies; the key generation activities conducted by the QKD link and physical layers are initiated, coordinated, scheduled, and terminated by the control plane. In the context of the data plane, \emph{data} refers to the key bits generated by the QKD devices over all the links in the network, which are independent of the end users' application traffic data transported in the integrated classical communication network. The QNL controls the lower layer activities according to requests from the KMS. The QNL consists of several functional areas that may be decomposed into a modular software structure, as shown in Figure~\ref{fig:qnl_functional_view} and described in Table~\ref{tab:qnl_functions_described}.

The control plane within the QNL establishes a path for the relay through by issuing key generation requests through a routing function. As part of its optimization, the routing takes into consideration the link capacities of the quantum network as well as the demand for key generation as determined through host demand monitoring by the KMS. The QNL establishes one or even multiple paths for key generation between any arbitrary pair of nodes, and performs automatic reassembly in the latter case. The QNL schedules key generation real-time in response to demand dynamics reported by the KMS. Its routing function is able to adapt to and perform computation on a dynamic network structure where nodes are usually fixed in place but not necessarily so; for instance, a satellite- or aircraft-based QKD setup could serve as a travelling node to provide temporary connectivity if a fibre link is unavailable or longer range is required. When multiple paths are employed, the key material can be combined at the destination site with some record-keeping overhead; for instance, the combination can occur through an exclusive~OR operation on the various key material inputs; various options are described in {\cite{salvail10}}. If some of the multiple paths may be unreliable, then Shamir's secret sharing scheme {\cite{shamir79}} can be employed to recover the key, but at the cost of some overhead in the transmission of key data along the multiple paths. All such techniques are compatible with our design.

The security policy may dictate which key forwarding tactics are allowed, and the KMS is charged with enforcing its ruleset by informing the network layer about any applicable policy. Such rules may be additional attributes of the defined security classes; in this case, the classes will dictate not only how key material is to be used, but also how it is to be generated. For instance, for higher security, the KMS can request the sole use of a direct quantum channel with the other site, if it is available. If forwarding along trusted nodes is allowed, then the KMS may specify whether a single or multiple concurrent paths are to be used, and how many hops along each path are acceptable. 

The data plane within the QNL temporarily stores the key material that is relayed between nodes; every trusted node along the path maintains a temporary local key pool for this purpose, as will be explained later in Section~\ref{subsec:qnl_data_plane}, and the QNL at the originating end performs final assembly of the key material. Once the QNL passes up the completed key bits to the KMS, the QNL no longer continues to store it. The KMS will maintain the key material within its own key pool until it is issued to users. Robustness is built into the fabric; if the QNL sees any service interruption, it notifies the KMS which performs a retry or passes the error up the stack, as appropriate.

\begin{table}
\begin{center}
\small
\begin{tabularx}{\textwidth}{|P{2.5 cm}|X|} 

\hline \multicolumn{2}{|P{12.5 cm}|}{\bf  The Control Plane suite in the QKD Network Layer coordinates all of the QNL nodes to form a QKD network and to generate raw key material according to requests from the KMS, and consists of the following modules:} \\
\hline Key Generation Messaging Function & This function enables both neighbouring and non-neighbouring QKD nodes to exchange key generation demands. \\
\hline QKD Routing Function & This function determines the path or multiple paths to be used for generating keys between any pair of nodes. \\
\hline  QKD Scheduling Function & This function schedules the task of relaying keys for different pairs of QKD nodes. \\

\hline \multicolumn{2}{|P{12.5 cm}|}{\bf The Data Plane suite in the QKD Network Layer generates raw key material through QKD for any arbitrary pair of nodes, according to instructions from the Control Plane, and consists of the following modules:} \\
\hline Key Relaying Function & This function performs the key relaying operations for a given pair of nodes, according to the schedule given by the QKD Scheduling Function. \\
\hline Temporary Key Pool Management Function & This function manages the local key pool to supply the key bits needed by the Key Relaying Function for other sites, as well as the ones needed by the local site. \\

\hline \multicolumn{2}{|P{12.5 cm}|}{\bf The external interfaces to the QNL, and the QNL modules that communicate via these interfaces, consist of the following:} \\
\hline Interface to the Service Layer & This interface provides a communication mechanism between the KMS and the QNL, through which the two layers can exchange key generation demand and response messages. The KMS can demand a particular operating mode for key generation, a generation rate, and amount of key material for particular sites, depending on the context. \\
\hline Interface to the QKD Link Layer & This interface provides a communication mechanism between the QNL and the QKD Link Layer (and ultimately the physical layer), through which key generation request and response messages are exchanged. This interface has to be extensible so that it can work with different QKD devices, as the command interfaces of those devices are typically vendor-dependent. In the context of the relevant QKD ETSI Industry Specification Group (ISG) standard {\cite{etsi2010}}, this interface could satisfy the ETSI QKD Application Interface API, which is used to obtain key material for an associated key handle. The network layer can specify Quality of Service (QoS) information to the link layer to request a particular bit rate, based on the demand data provided by the KMS. \\

\hline
\end{tabularx}
\caption{Descriptions of the high-level functions of the QKD Network Layer.}
\label{tab:qnl_functions_described}
\end{center}
\end{table}

The QNL will ultimately optimize the key generation across the entire network so that throughput in the system is maximized. For users that are of higher priority, the policy may typically dictate frequent session key refresh and perhaps even a one-time pad for very limited use for certain users. In this case, these key generation demands may be weighed more heavily in the optimization calculation. 

\subsection{The QKD Link Layer}

Immediately below the QNL, the QKD Link Layer (QLL) produces raw quantum key bits over each link by executing a QKD protocol. It establishes quantum key material between connected node pairs and provides it to the QNL. There is a plethora of QKD protocols and link technologies to choose from, with varying implementation complexities, key rates, and robustness; some are mentioned in \cite{padamvathi16} and \cite{lo15}. A ``quantum Internet'' will likely need to integrate various quantum technologies \cite{pirandola16}. Metropolitan-scale distances are achievable through a combination of fibre and limited free-space links. Longer distances could be supported by quantum repeaters and satellites, and physical routing accomplished via optical switching, whether active or passive. Resource sharing is possible with the multiplexing of QKD and classical signals. Ultimately, the target context will help select suitable QKD technologies.

The QLL can expose switching and addressing functionality to the QNL, so that the QNL can configure QKD devices and their connectivity to achieve an effective setup for the routing of key generation. However, only the QLL has direct access to the quantum link; any instructions issued from the QNL must traverse through the QLL layer. The QLL can make use of dedicated infrastructure as well as shared resources to carry out the key generation. The QLL also informs the QNL of its current rate as well as its practical limits, and additional constraints, so that the link capacity can be understood; for example, there may be a predicted temporal window of opportunity for key generation via free-space transmission. 

\section{High-Level Procedures in the Communications System}
\label{sec:procedures}

A number of critical procedures are always executing in the QKD communications system, some of which will result in interactions spanning multiple layers in the proposed software architecture. The procedures are described at high-level in the following subsections.

\subsection{Session Key Issuance Within the Host Layer}
\label{subsec:key_issuance}

A generic protocol to initiate and conduct secure communication between computing hosts, based on the use of quantum-generated encryption keys, is described. The protocol is host-based; it provides full control of the communication session to the hosts. Although this particular protocol is formulated from scratch and requires custom implementation work, it is instructive in its unique approach to session key agreement based on the use of QKD. Direct integrations of similar mechanisms with existing standardized transport protocols such as TLS have also been investigated. For brevity, the generic protocol will be examined at this point: suppose that Alice, the originating host on Site~A, wishes to initiate a secure communication session with Bob, a host on remote Site~B, and requires confidentiality of user data transmitted over the network. Alice and Bob may each communicate with their respective local KMS via their LAN network or a separate encrypted channel. Prior to the session initiation, or alternatively as a direct result of it, Sites~A~and~B will have engaged in the QKD protocol so that they will share a pool of key material. In this protocol, the hosts are responsible for completing the key negotiation protocol sequence by communicating with their respective KMS's, and between the hosts themselves. As multiple communication sessions may be initiated between each pair of hosts in the metropolitan network, each session will be identified by a globally unique session identifier. The pair will initially require mutual authentication prior to key negotiation, which may be accomplished through the use of a pre-shared secret or public key signatures (that are not quantum-safe but are used only in the short-term, and issued by a certificate authority) \cite{etsi2015}. Alice commences by requesting a new session key from her KMS. Alice identifies that the remote host resides on the remote site B through a Domain Name System (DNS) or similar query. Alice assigns a globally unique session ID to the new communication session being initiated. Alice's KMS constructs a new session key by obtaining quantum key material from the quantum key pool, if it is not already immediately available for assignment. Alice's KMS must enforce any security policies that are defined. Alice's KMS obtains advice from the policy engine based on a previously injected policy and the identities of the parties involved. The policy engine returns pertinent parameters enforced by the policy, such as the minimum key length and maximum allowed session key lifetime.

The KMS provides a symmetric session key to Alice constructed from the material in the quantum key pool; the key may be a block cipher key. Alice caches it for use in secure communication with Bob for the duration of the session and no longer than its required lifetime. The KMS also provides an associated key selection information packet sufficient for the remote host to obtain the same key on the remote site. This packet is encrypted using an existing inter-site symmetric key to secure communication between the pair of KMS's on separate sites. The KMS also provides a Message Authentication Code (MAC) for the packet to detect and prevent tampering. Alice then provides the packet to Bob so that he can recover the same session key from his own KMS. Bob's KMS verifies what constraints must be enforced on session key generation and management, based on an injected policy and the parameters specified by the requesting host; modification or re-negotiation of some parameters may occur. Bob then sends back a confirmation to Alice, and secure communication between the two hosts can then begin.

During construction of the session keys from the quantum key material in the pool, it is possible for a race condition to occur as a result of the possibility of multiple hosts simultaneously initiating communication sessions, and thus session key requests, at each site. The likelihood of this occurrence increases with the size of the host population and the frequency of their communication with hosts on external sites. For instance, Alice may initiate a session request and be granted a session key from her KMS. Before Alice sends a session negotiation message to Bob with the key selection information, another user Charlie on the remote site~B may engage his KMS. Charlie may make a similar session key request and be granted the same session key on his end that Alice intended for Bob. In this case, Bob's KMS will report an error to Bob which will get relayed back to Alice and result in the session key request having to be re-tried with a new key. In a highly scalable system, such race conditions may occur at significant frequency and result in unnecessary and frustrating delays for users. To reduce the impact of round-trip error handling, an appropriate key allocation strategy may be utilized that lessens the chance of race conditions occurring. For instance, although the quantum key pools between sites A~and~B are synchronized, Alice's KMS can start consuming key material from the beginning of the pool, and Bob's KMS can start consuming it from the end. However, an effort should be made to reduce fragmentation of the key pool, since new key material will be generated in the same memory or disk region from which it is consumed; such fragmentation could lead to wasted space. A possible tactic to mitigate fragmentation is to designate a small portion of the quantum key pool as a working set, and have each KMS consume key material from a different end within this working set only.

Care should be exercised when independently issuing copies of key material produced by QKD to multiple parties across sites to avoid security issues. For example, if Alice is issued a session key that Charlie (as a third-party) obtains on the remote site for an unrelated session rather than Bob (Alice's intended recipient), then none of these parties must be allowed to proceed with the use of the keys and risk the possibility of Alice and Charlie reading each other's communication. The KMS must deny a host from using key selection information to retrieve a key previously issued on the same site; furthermore, a host being granted a key must receive confirmation from the remote host that the same key was successfully retrieved, before the key in question is used to encrypt user communication. There are ways to enforce this problem beyond host protocol rules, such as fully synchronizing the KMS's on each key issuance and updating their key pool status, or by maintaining separate key pools for each host at each site; however, these techniques carry overhead and scalability may be adversely affected. Another approach may be to use a token system: when Alice requests a key, she obtains a token, not the key itself yet. The token is then sent to Bob to retrieve the actual key on his end. Upon confirmation of this, Alice can then use the token to get her own copy of the key. In this way, the keys are not immediately issued by the KMS's until there is assurance of no race condition being present, whether intentional or not. Ultimately, this issue must be accepted, with hosts being trusted to obey the protocol, or a suitable solution must be devised.

\subsection{Integration with Standard Security Protocols Within the Host and Third-Party Local KMS Layers}
\label{subsec:standard_protocol_integration}

In addition to defining generic protocols, the process of agreement of QKD-generated keys has been integrated by using the pre-shared key cipher-suites of TLS (Transport Layer Security) \cite{rfc4279, rfc5246}, IPSec \cite{rfc2409}, and Kerberos \cite{rfc4120}, where the key selection information is transmitted using supported fields. Generically, \emph{pre-shared key} is a secret that was previously shared over a secure channel; in this case, QKD-generated key material becomes a pre-shared key that enables a security protocol and may even be an input to further key generation algorithms. The integration with the standard cipher-suites requires minimal changes to the client implementation and avoids additional complexity, which is enemy to practical security engineering as explained in \cite{ferguson10}. The use of QKD key material results in a quantum-safe handshake and key agreement; an appropriate cryptographic suite and key length are then negotiated to provide quantum-safe computational security for encrypted user data. Information-theoretic security through a one-time-pad is not directly available in standard deployments unless a custom cipher suite is made available that consumes quantum key material to encrypt user data bit-by-bit in a one-time pad manner. The high-level approaches to the integrations are as follows:

\begin{itemize}
\item In the case of TLS, security is provided at the transport layer. The pre-shared key (PSK) identity hint field can be re-purposed to contain the quantum key selection information encoded as a string of numeric characters, and encrypted by an inter-site key. It will be transmitted in the client key exchange message between the originating and the remote hosts, and supplied to the respective KMS's in key grant messages, in order for the two hosts to retrieve the final session key. To provide computational security while being considered quantum-safe, when using a QKD-generated key as the pre-shared key, AES-256 can be used for encryption and HMAC-SHA384 for message integrity and authentication \cite{etsi2015}. 
\item In the IPSec protocol suite, the Internet Key Exchange (IKEv1) protocol is used to set up a security association (SA) to provide security at the IP packet layer. In the less-verbose Aggressive Mode, the key selection information can be supplied in the identification payload between the initiator and the responder hosts; this is in contrast to the original purpose of the payload being to create a Diffie-Hellman shared secret which would not be quantum-safe \cite{etsi2015}. This PSK is typically used for mutual authentication in the first phase of the protocol in order to set up a secure channel for key exchange. The authentication key, or the shared master secret, is derived from the PSK through IPSec's key construction strategy. In later phases, the IPSec tunnel is established and encryption keys may be generated. 
\item Finally, Kerberos provides a mechanism for mutual authentication of a client and server located on different domains. Each party retrieves the same session key from their respective Ticket-Granting Service (TGS), which is derived from key material generated through QKD between these local and remote services. The key selection information can be specified in the service ticket issued by the local TGS and provided in encrypted form to the remote server (located on another site); from it, the remote TGS will then provide the final session key to the remote server to encrypt user communication. Kerberos is widely used to provide Single-Sign On (SSO) functionality by having the client provide service tickets, that confirm the client's identity, to various sites without the necessity of re-entering credentials. With the described integration of QKD, each service ticket could contain a reference to a different session key from the quantum key pool accessed by the TGS, so that each client-server communication is protected.
\end{itemize}

The integration of each protocol provides a different but secure avenue through which the quantum key selection information is provided as a pre-shared secret to both parties, and then used to construct an encryption key to provide confidentiality for the exchange of user data, or to provide an authentication function, depending on the context; all of this is accomplished without significant modification to the protocol. In the case of the TLS and IPSec protocols, which are essentially consumers of the QKD-generated key material, their client implementations are assumed to occupy the Host Layer in the architecture of the proposed QKD system. In the case of Kerberos, client logic may exist within the Host Layer, while components such as the TGS would be expected to lie within an additional Third-Party Local KMS Layer. In all cases, the next layer down is the KMS Layer of the QKD system, which supplies the QKD-generated session keys to be used by the protocols.

Adapting to multiple standard protocols is useful, as although they are highly versatile and complex, they are generally used for different purposes. IPSec is embedded within the Internet Protocol (IP) stack and is typically used to create secure Virtual Private Networks (VPNs) that operate at the network layer of the TCP/IP model, so it protects all upper-layer transports so that remote hosts function as if they were directly attached to the local network; however, it requires specialized client software and supporting network devices. TLS functions above the Transmission Control Protocol (TCP) layer, can be utilized by lightweight clients, and is supported by all browsers for use in e-commerce; it is used for server-side authentication through certificates and provides remote access, fine-grained access controls, and end-to-end data encryption for applications. The Kerberos protocol is generally used for mutual authentication of a client and server through a third-party with the use of passwords; it enforces user access rights to multiple services and permits single-sign on, as described above; however, the Kerberos server can be a single point of failure. 

\subsection{Demand Monitoring and Key Generation Requests Within the Service Layer}

When requesting initiation of QKD key generation by the QNL, the KMS specifies one of the following types of key generation to conduct:

\begin{enumerate}
\item \emph{Continuous} key generation at a particular rate with a given remote site. Key material is generated while the key pool has capacity remaining. Once it fills up, key generation can be temporarily suspended, or existing content in the pool can be overwritten. The KMS may simply request the maximum rate supported by the current QKD equipment and infrastructure that are installed; these are reported to the QNL by the link layer.
\item A \emph{one-time} demand for generating key bits of a specified amount for a given remote site. This may occur when a client requests a new session key for the first communication with a remote site since the last time that the QKD process ran, and the pool is otherwise empty. The quantity of key material to generate is based on the expected demand and may be periodically adjusted. It should be limited by the size of the database that has been allocated for the pool.
\end{enumerate}

The KMS will request key generation according to the current operating context. The various heuristics that will affect this choice, particularly for the continuous generation mode, include the following:

\begin{itemize}
\item The expected steady-state demand for key material will dictate the continuous key generation rate; the demand is based on the expected number of hosts at a site, and the expected frequency at which they will request new communication sessions at particular times. In turn, based on the resulting required rate of session key assignment and key sizes involved, the key generation rate can be estimated. The frequency can be estimated based on historical traffic patterns such as host arrival rates. 
\item The expected peak demand for key material at regular or irregular times can determine the amount of key material that needs to be cached by the KMS. The ability to satisfy peak demand can be accomplished by specifying a statically-allocated key pool database that is sufficiently large, or by allowing it to grow to a sufficiently large size. Once the database is exhausted during the peak period, hosts will have to wait for new key material to be generated.
\item The key assignment strategy will also determine the key generation rate. The quantum key material may be used to construct session keys or it may be used as a one-time pad, as dictated by the security policy in place. The key size to employ will also be specified in the policy. In the case of session keys, one key will be assigned for one entire communication session, and the key generation rate will set based on the number of hosts and the number of sessions that they initiate. In the case of a one-time pad, one bit of key material will be consumed for every bit of user data, so the rate will be set based on the total amount of data to be transmitted by a site. The highest security level will typically require the implementation of a one-time pad.
\item The frequency at which session key material needs to be refreshed will affect the demand and thus the key generation rate for the same expected number of hosts. The frequency of key refresh is specified by the security policy. Key refresh may only be required at higher security levels. 
\item The economical costs of the network channels used to conduct the QKD protocol may limit the key generation rate that is employed in practice. Likewise, the infrastructure utilized for QKD will impose a practical upper bound on the rate. If the costs limit the key generation rate to a level that cannot meet the expected demand, the security policy may need to be sufficiently relaxed to allow the re-use of key material; this may be manifested by key derivation algorithms.
\end{itemize}

The key generation rate may be dynamically adjusted by a KMS while QKD is in operation. The request for key material from the KMS may be impossible to completely fulfill by the QKD Network Layer. The physical limitations of the QKD hardware and supporting infrastructure that are installed will impose an upper bound on the key generation rate that can be implemented. Limitations will apply as a result of the kind of technology employed, the network infrastructure in use, and the physical length of the quantum channel between sites. As only the QNL has knowledge of the underlying infrastructure, it will limit the KMS requests as appropriate based on the QKD link capacities reported up from the QKD Link Layer.

The KMS may alternate between the two operating modes to best serve its local host demand for session keys, which it is constantly monitoring through various heuristics. The quantum key pool that it maintains can serve as a cache for a surge in traffic. Once the pool is depleted, the KMS can request on-demand key generation by the QNL. On the other hand, if there is a constant level of demand, the KMS can request continuous key generation. A hybrid operating mode is also possible, where a certain rate of key generation occurs in continuous mode, and reserve capacity is left for occasional use in the on-demand mode; this strategy may be appropriate if utilization of existing quantum links carries a cost, such as on shared fibre-optic lines. The proposed design is flexible enough to allow the operator or owner of the integrated system to pursue and achieve different business goals.

Multiple security classes of connections may be defined within the security policy managed by the policy engine, which specify how the quantum key material is to be utilized to provide data encryption. The levels are shown in Table~\ref{tab:security_classes}.

\begin{table}
\begin{center}
\small
\begin{tabular}{|C{1.5cm}|P{10.5cm}|} 
\hline \bf Security Class & \bf Strategy \\
\hline \hline \makecell[t]{5 \\ (highest)} & The quantum key material will be utilized as a one-time pad. The same amount of key material is required as the size of the message being encrypted; this use will consume the quantum key material at the highest rate. \\
\hline \makecell[t]{4 \\ (default)} & The quantum key material will be utilized to generate session key material. Each new secure connection initiated by a host to a remote host will be assigned a new single session key. The session key will expire after a specified interval, and will need to be refreshed. \\
\hline 3 & As in class 4, but no key refresh occurs. \\
\hline 2 & As in class 3, but if insufficient new quantum key material is currently available, then key extension will occur to generate a session key from the current material. A suitable key expansion technique may also be utilized, such as the Rijndael key schedule in AES \cite{fips197}. This scenario may occur in a highly scalable application, where the number of users engaging in simultaneous communication with other sites may exceed the maximum quantum key generation rate that may be attained given hardware and infrastructure limitations. \\
\hline 1 & The quantum key material is used to generate a host-specific key that is used for all communication sessions that the host engages in. \\
\hline \makecell[t]{0 \\ (lowest)} & Insufficient quantum material exists to provide a key, or QKD key generation is not functioning for some reason. A classical key generation mechanism will be utilized, instead, to produce a session key; the mechanism must be quantum-resistant, if possible, and key refresh should occur. This class may be utilized if the QKD protocol cannot  be currently operated, or if the key generation rate cannot keep up with the current demand for key material and hosts cannot be delayed. \\
\hline
\end{tabular}
\caption{Descriptions of the various classes of security levels in the security policy.}
\label{tab:security_classes}
\end{center}
\end{table}

\subsection{Quantum Key Pool Management Within the Service Layer}

\begin{figure}[htbp]
\begin{center}
\includegraphics[scale=0.40]{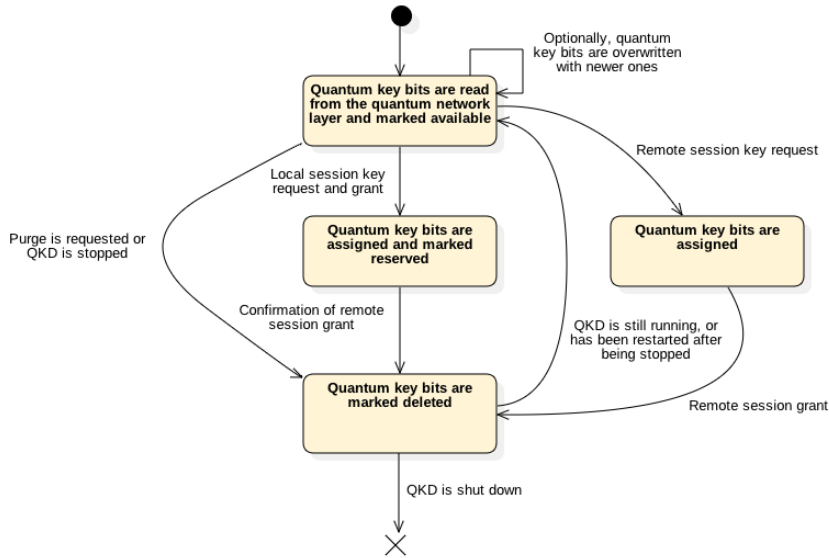}
\end{center}
\caption{A state diagram shows the functional states of the Quantum Key Pool contained within the KMS and the possible transitions between them. QKD-generated key bits are assigned from the quantum key pool during key grant to a local host, and then reserved until the same bits are retrieved by the remote host, so that the same bits cannot be assigned to more than one host. If the key bits that the remote host tries to retrieve have already been assigned to some other host, due to a race condition, then an error is reported and the entire process must be restarted.}
\label{fig:qkp_states}
\end{figure}

The quantum (QKD-generated) key bits occupying the quantum key pool undergo a series of state transitions during normal operation of the system, as show in Figure~\ref{fig:qkp_states}. When the key pool is initialized, quantum key material is read from the quantum network layer and injected into the key pool; this is effectively a copy to a permanent store. When a local host makes a request for a new session key in the context of a local session key grant, a subset of the bits is assigned as a session key, and reserved. When the remote host is contacted and obtains key selection information and retrieves the same key material on the remote site, in the context of a remote session key grant, the key bits are immediately purged on confirmation of the key grant. The key bits are then replenished with new ones as a result of the QKD key generation process. Note that the key material cannot be deleted from the originating site until it is assigned on the remote site; the reason is that the contents of both quantum key pools must be exactly the same at all times; this guarantees that the remote host is able to obtain the same session key as the local host through a remote session key grant. This principle also applies to the state associated with each key bit. 

Normally, quantum key pool bits that are available for assignment are preserved until the assignment occurs. Once the quantum key pool becomes full, then no new key bits are read. However, if the QKD process is operating in the continuous mode, then it is also possible for the available bits to be overwritten with new ones. It is also possible that once the pool becomes full, then additional space may be allocated to increase the size of the pool in order to contain more key material. The remote site must be informed that it needs to allocate more space, too. Note that special conditions can occur, such as: a system shutdown (due to an error or disaster or planned maintenance outage), interruption of the QKD process, or unexpected loss of synchronization of quantum key pool content. These conditions will cause the key pool contents to be purged immediately. The QKD protocol can be restarted once the system returns to a normal state. Backup and restore functions can reduce the recovery time if a secure backup store exists. Synchronization of the key pools belonging to a pair of sites occurs through direct communication of the respective KMS layers through the KMS peer interface.

The key pools on both sides of a quantum network connection must match in content so that the same quantum key material can be referenced and discovered during session key negotiation conducted by the site hosts. If content synchronization is lost for any reason, then the key pools will be purged, and new quantum key material will be generated to regain a known good state. A policy may enforce the expiration of the pool content after some time has elapsed or a threshold number of session keys have been assigned, which will also lead to key re-generation. Such re-generation may be automatic or will occur as the result of normal operation of the QKD protocol.

\subsection{Key Generation, Routing, and Scheduling Within the Control Plane of the Network Layer}

The Key Generation Messaging Function (KGMF) in the QKD Network Layer enables both neighbouring and non-neighbouring QKD nodes to exchange key generation demands. The local KGMF software module within the QNL receives key generation requests from the KMS with either a continuous or one-time demand operating mode specified. Consequently, the KGMF will form a Key Generation Message (KGM). For continuous key generation, this message will be broadcasted to the whole trusted network so that every node in the QNL formulates the same map of global demand, which is critical to ensure the correct operation of the routing function. The KGM will be routed over a transport available to the QNL, such as a hop-by-hop TCP connection. The messages over a given TCP connection will be encrypted by utilizing a site-to-site key. 

For on-demand key generation, the KGMF will pass the information of the remote site and the length of the key requested to the scheduling function, which will place this task in the appropriate queue for scheduling. Occasionally, the KGMF may also receive from the KMS a command terminating the key generation for a remote node. This command will trigger the KGMF to broadcast a stop message to the whole network; every node receiving it will update its copy of the global demand matrix, and will trigger the QKD Routing Function (QRF) to take action accordingly. The QRF determines the path or paths to be used for generating a certain amount of, or a sustained rate of, key bits between any given pair of nodes. 

The QRF has an initialization phase and a normal working phase. In the initialization phase, each node needs to collect the following two inputs:

\begin{enumerate}
\item From the lower-layer QLL, the \emph{key generation capacities} over all quantum links in the network, and the identities of the neighbouring nodes associated with the links.
\item From the upper-layer KMS Layer, the \emph{key generation demands}, which depend on the system operation mode. If the system is operating in \emph{continuous mode}, a node collects key generation demands between itself and each of the other reachable remote nodes in the network, as well as those of each and every pair of other nodes. An intermediary relay node will record the requested key generation rate for the source and the destination node pair in each and every KGM that it routes. If the system is operating in \emph{on-demand mode}, the QRF collects key generation demands locally only. 
\end{enumerate}

Each node employs the principle of \textit{link-state routing} \cite{tanenbaum2011} to determine the best key generation paths in the network. Each node exchanges custom Link-State Advertisements (LSAs) with all of its neighbouring nodes, adapted from those found in the Open Shortest Path First (OSPF) TCP/IP Internet routing protocol that uses a Link State Routing (LSR) algorithm \cite{rfc2328}. An LSA communicates a node's local routing topology to all other local nodes in the same area by flooding. An advertisement packet includes the identity of the node itself, the identities of all the other nodes to which it is directly connected, and the associated link capacities in terms of the key generation rate; it also includes a sequence number that is necessary for the recipients to keep track of the latest information on the link status. Each node, after receiving an LSA with a newer sequence number, incrementally constructs a graph of the connectivity to the network; after receiving a complete set of LSAs, it will be able to compute the whole network topology. From it, the node can then independently calculate the current best logical path for key generation to every possible endpoint in the network. Each collection of best paths will be used to produce the node's local routing table by employing a scheme such as shortest-path or least-cost-routing computation. 

In the normal working phase, the QRF's main tasks are determined by the operating mode of the system:

\begin{itemize}
\item In the \emph{continuous mode}, the QRF solves the problem of maximizing the key generation workloads, referred to as \emph{flows}, within the network; it strives to maximize the overall proportion of requested demand that it can satisfy along all paths, such that each flow is bounded by the capacity of the quantum link. After the solution has been obtained, the QRF will then know the key generation workload for any QKD link; this information will be passed on to the scheduling algorithm. 
\item In the \emph{on-demand mode}, the QRF finds the route from the source node to the destination node for a given demand. Consulting the routing table produced in the continuous mode of operation (and updated when the topology changes), if it exists, each node can route all the key generation messages received. Occasionally, the QRF will need to update the network topology and recalculate the workloads. This will be triggered by events such as node or link failures, which can be reflected in new LSAs from directly-affected neighbouring nodes.
\end{itemize}

The routing capability of the QNL takes into consideration the residual capacity of all the QKD links within the whole network. In link state routing, each node can independently calculate a new route in case of a node or link failure. All nodes need to have a highly-consistent or synchronized view of the topology; this is feasible because the number of sites in the network is relatively small, topology changes due to planned new nodes joining or existing nodes leaving the network are infrequent, and the sites of the integrated network should be designed to operate with high reliability; such changes will be made known to all the nodes in the network rapidly via the "distributing maps" stage of the link-state routing protocol \cite{kurose13}. All nodes must have the capability of detecting changes in the connectivity between themselves and their neighbours; there are at least two ways to achieve this: firstly, the underlying QLL will send error indication messages to the QNL if it detects any error at the link layer that changes the capacity of the link; secondly, changes can be indicated in the ``hello" messages exchanged between neighbours \cite{kurose13}. If all retries fail, then an updated LSA will be sent out to announce the new topology to other nodes in the network.

In the continuous mode of operation, the demands are expressed in terms of key generation rates for source-destination pairs. The problem can be formulated mathematically as a Maximum Concurrent Multi-Commodity Flow Problem (MCFP) \cite{girard90}; such a problem may be solved using a fast approximation algorithm such as the one proposed in \cite{karakostas08}. The organization that runs the network can learn from the operational history and gradually adjust the demands to achieve a trade-off between maximizing the utilization of the network and the unbalanced demands of different source-destination pairs. Another way to deal with the problem is to formulate it as a Maximum Multi-Commodity Flow Problem, where the objective is to maximize the total amount of flows through the network or to maximize the network utilization, as in \cite{fleischer00}. The solution is an important input to both the control and data planes. Special treatment can be applied to higher priority demands within source-destination node pairs; one way to cater to such a need is to pre-allocate QKD capacities along chosen paths for such pairs in the network, and subtract these capacities before solving for the general capacity problem.
  
The QKD Scheduling Function (QSF) is responsible for scheduling the tasks of relaying QKD key bits for different pairs of QKD nodes. The output of the scheduling algorithm in the QSF is a sequence of work tickets on a per-link basis. When a local node has more than one immediate neighbouring node, the algorithm handles all workloads over each individual link independently. 

When the system is operating in the \emph{continuous mode}, the QSF inputs the workload assignments from the QRF. It then schedules the tasks for any given link following a deficit-weighted round-robin algorithm \cite{shreedhar95}. When the system is operating in the \emph{on-demand mode}, the scheduling algorithm will follow a simple First-In-First-Out (FIFO) queuing mechanism. That is, all the flows over one outgoing link will be put in a single queue for that link, and will be served according to their arrival sequence.

\subsection{Key Relaying Within the Data Plane of the Network Layer}
\label{subsec:qnl_data_plane}

\begin{figure}[htbp]
\begin{center}
\includegraphics[scale=0.50]{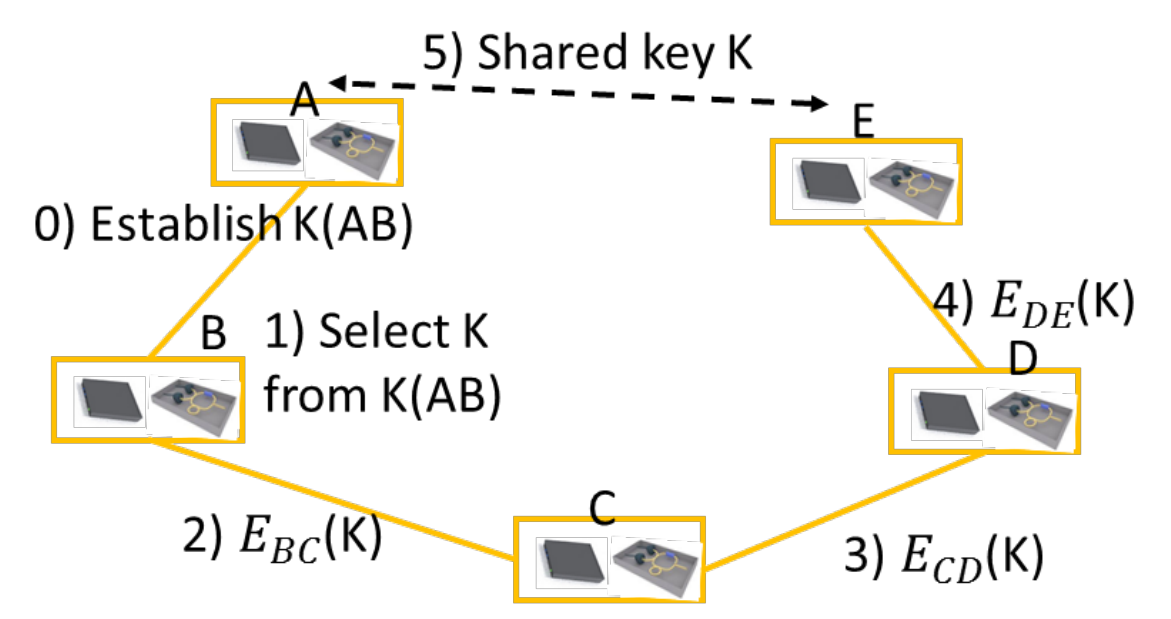}
\end{center}
\centering
\caption{A possible mechanism for key generation and relay across a trusted node network is illustrated. Suppose that a host on site~$A$ wants to establish a secure communication session with a host on another site~$E$. The two parties need to establish a shared key $K$ as their session key. As the two sites are not directly connected via a quantum link, the Quantum Network Layer (QNL) will find a path from $A$ to $E$ that traverses over trusted nodes $B$ to $D$. In the first step 0, part of the initialization process, $A$ and $B$ will establish a key stream $K_{AB}$ over their quantum link using QKD, and then in step 1, $B$ will select a subset of this stream as the session key $K$ to be used by endpoints $A$ and $E$. In step 2, $B$ and $C$ will establish a key stream $K(BC)$, and so on in the following steps for each pair of trusted nodes until $E$ is reached. The session key $K$ will be ferried across the chain of nodes by being encrypted by the additional key streams and transmitted over a conventional network between each pair of trusted nodes; for instance, re-visiting step 2, after establishing $K(BC)$, $B$ will send the session key $K$ to $C$ such that it is encrypted with the keystream $K(BC)$, serving as a one-time-pad, as denoted by $E_{BC}(K)$. After $C$ receives the encrypted key, the key stream $K(BC)$ is discarded. Thus, each node in succession decrypts the session key and forwards it on to the next node in secure fashion until the endpoint is reached, and all key streams used for the encryption will be deleted, as the one-time-pad forbids re-use. At the end of the process, nodes~$A$ and $E$ will share the same session key~$K$ and use it to secure communication between themselves.}
\label{fig:trusted_node}
\end{figure}

The Key Relaying Function (KRF) is responsible for conducting QKD key generation within the network of quantum links. It generates and relays key bits across trusted nodes in a mechanism like that described in \cite{salvail09}. The KRF processes the flows over any given quantum link as scheduled by the QSA based on the flow assignment dictated by the QRF. As a prerequisite, all links along a path connecting source-destination endpoints must have established their own key pools for secure communication between the neighbouring nodes. Referring to Figure~\ref{fig:trusted_node}, suppose that node~$A$ is the source node and node $B$ is the immediate downstream neighbour in the chosen path to destination node~$E$. Node~$B$ will select bits from its local key pool shared with~$A$ as the key~$K$ to be eventually passed to node~$E$. Node~$B$ encrypts the key~$K$ using a one-time pad (OTP) with shared key bits between itself and the next downstream node~$C$. Node~$B$ forwards the encrypted key to node~$C$, and so on, until the destination node~$E$ is reached.

The KRF at each node performs the relaying activity following the work ticket output from the scheduling algorithm in the QSF. For each ticket, the KRF retrieves the corresponding decrypted key material from the local temporary key pool. If the destination is the local node to which it is attached, it passes the key material up to the KMS. Otherwise, the KRF retrieves temporary key bits of length equal to that indicated in the ticket, which were previously generated through the QKD protocol. The KRF then performs the OTP operation (possibly by combining through an exclusive-OR the temporary key bits with the to-be-relayed key material) and sends the result to a neighbouring site. The transport channel over which these key relaying events occur is dedicated for the data plane traffic of the QKD network. This persistent connection between neighbouring nodes can be realized through a classical communication channel between neighbouring QKD devices and accessed at the QLL level; it may be authenticated by using a site-to-site authentication key. 

Improvements in the design of the QKD network topology can yield greater performance in key generation; for instance, backbone nodes can provide access to a higher-capacity sub-network. Such high-capacity data plane channels can be incorporated into the workload assignment problems.

The Temporary Key Pool Management (TKPM) function temporarily stores and manages key material at the QNL layer so that the Key Relaying Function can make use of raw key material as the input into the one-time pad for key encryption. The raw key material generated falls into three categories, which are allocated separate buffers within the temporary key pool:

\begin{enumerate}
\item The key bits that will be used by the host traffic between the local site and a directly-connected remote site.
\item The key bits that will be used by relaying the keys to be used for host traffic between the local site and a remote site via one or more intermediary nodes.
\item The key bits that will be used by relaying the keys for other node pairs that are using the local site as an intermediary node. 
\end{enumerate}

\begin{figure}[htbp]
\begin{center}
\includegraphics[scale=0.60]{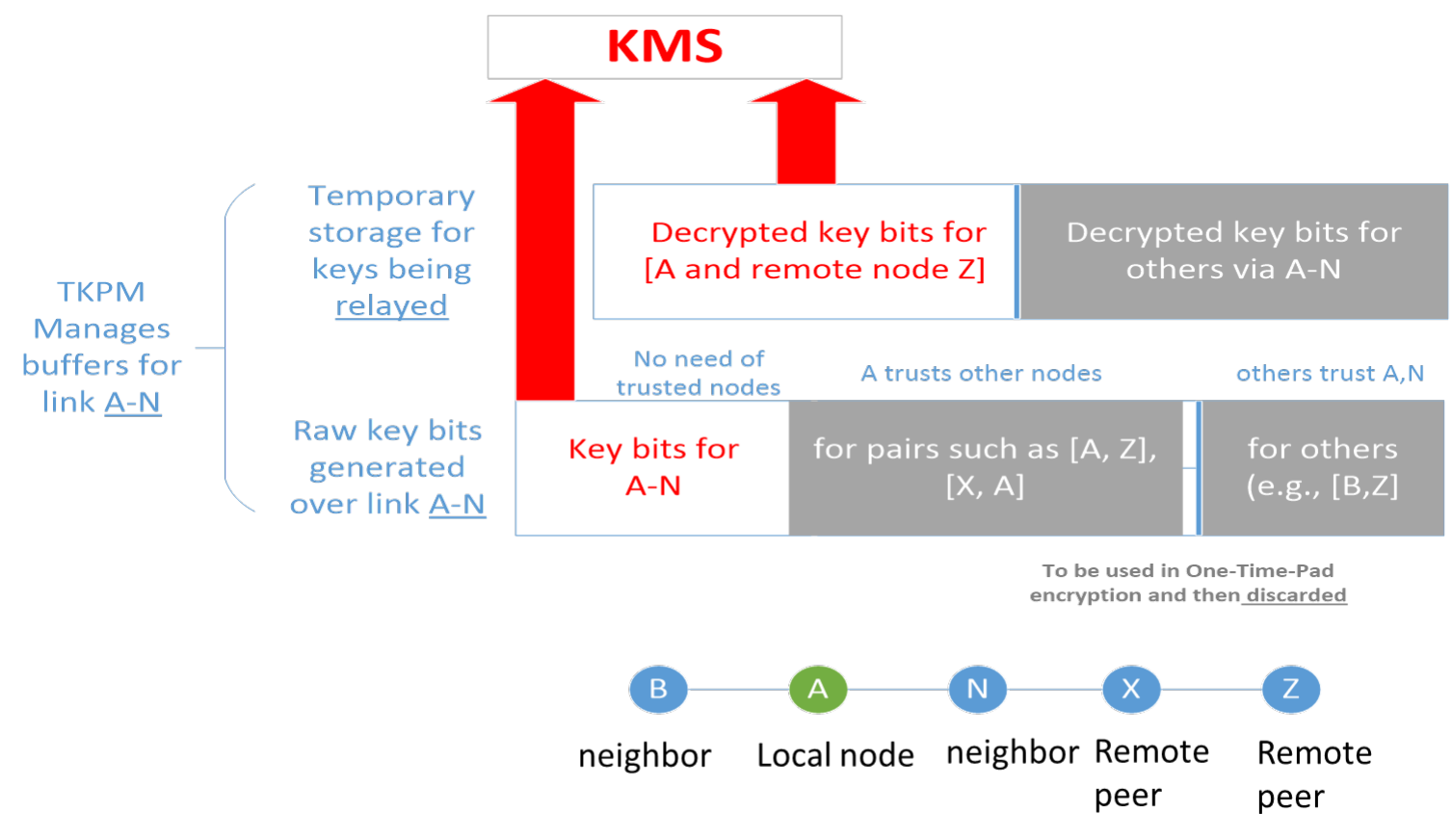}
\end{center}
\centering
\caption{Quantum key material buffers, also called temporary key pools, are allocated for each site to fulfill the temporary key pool management responsibility of the data plane in the Quantum Network Layer. Various buffers are required for the various categories of key bits that are generated. Refer first to Figure~\ref{fig:trusted_node} for an explanation of how key generation occurs via trusted nodes. Now refer to the example node connection graph illustrated here: between two nodes that share a direct quantum link (such as $A$-to-$N$, where the endpoints for secure communication are $A$ and $N$), the raw key bits output from execution of the QKD protocol are immediately provided to the key pool in the KMS Layer above. If $A$ is an endpoint node without a direct quantum link to the other endpoint (such as $A$-to-$Z$), then $A$ must trust other nodes ($N$ and $X$ in this case) to form a relay, so it allocates a temporary buffer for the key stream with adjacent node $N$ from which the session key is extracted. If $A$ is a trusted node intermediary between two other endpoints (such as $B$-to-$Z$), then it will keep a buffer for the key streams established through QKD with its immediately adjacent nodes ($B$ and $N$), which are used in the encryption of the session key being transported to the remote endpoint~$Z$. Buffers will also be maintained for session keys that are being relayed. For instance, if $A$ is an endpoint (such as in $A$-to-$Z$), it will pass session keys that are relayed to it from the remote endpoint~$Z$ up to the KMS; $A$ will also maintain a temporary buffer for a session key that it decrypts and then re-encrypts as part of the relay to other endpoints. Raw quantum key material is discarded once the relay function is fulfilled, and the material is passed up to the KMS Layer as required, to reclaim storage and minimize risk.}
\label{fig:temp_key_pool_management}
\end{figure}

The TKPM also manages the temporary storage of the keys being relayed by the local node, which is used by the Key Relaying Function. Figure~\ref{fig:temp_key_pool_management} shows the different categories of key bits being managed by the TKPM for various possible links.

\section{Trust and Security in the Communications System}
\label{sec:trust_and_security}
 
\subsection{Network Trust Model}

The KMS is entrusted by all hosts within each site to aid in securing host communication. In particular, the KMS will generate, manage, and distribute symmetric private key information used to encrypt host communication. While the KMS is able to decrypt host communication, hosts will not share their communication content directly with the KMS.

Any intermediate trusted nodes that may be used to conduct the key generation and relay processes may have sufficient knowledge of the quantum-based key material to construct decryption keys and access user communication. In fact, the key relay protocol requires decryption and re-encryption of key material at each intermediate hop. Thus, trusted nodes participating in the QKD process must be fully trusted by both origin and destination sites.  An alternative would be to employ multiple paths for the relaying of quantum key material between the source and destination sites and through the trusted node network. As long as the nodes along at least one path are not successfully compromised by an attacker, then the security of the key material is kept intact. An attack could be mounted by an active adversary or simply by having the trusted nodes function in an honest-but-curious manner such that they occasionally eavesdrop. If all the key bits generated from multiple paths are combined to construct the final key, then single nodes along different paths will have access to just subsets of the entire key material, while the source and destination sites have access to all of it. A successful attack on one path will not compromise the key material generated on another path. However, the overall effective key length may be reduced in proportion to the subset of key material that a single node, compromised by an attack, is forwarding along its own path. To maintain the same effective key length, the use of multiple paths will require an increase in the transmission of key bits (e.g. a doubling of keys bits when using a pair of key relaying paths, compared to a single path).

\subsection{Authentication of Sites}
\label{sec:authentication_sites}

Sites engaging in secure communication, for instance to exchange control messaging traffic, must be authenticated so that they can trust each other in multiple scenarios and across multiple layers of the communications stack. For instance, sites must communicate with each other securely at the KMS layer in order to coordinate the start and termination of the QKD protocol in the network layer below. They must also communicate at the network layer in order to relay the quantum key material over intermediate hops. In all cases, one or more TCP channels may be typically utilized; because the channels may be public, confidentiality and authentication is required through the use of an inter-site, or site-to-site, key; one of the following means may be used as the basis for the key, with the first option being 
simplest in the early stages of QKD networks, but other alternatives may be used instead as the networks become more sophisticated:

\begin{enumerate}
\item A pre-shared key is utilized. The first instance of this key is generated through some means other than QKD (which is not yet executing), and then disseminated through a secure out-of-band channel. It is then re-assigned during initial operation of the QKD hardware, using QKD-generated key bits. For instance, it may comprise the sequence of bits found in a reserved area of the quantum key pool, once those bits are generated, and then refreshed regularly with new key bits.
\item Short-term public cryptography is utilized as a boot-strap to initially authenticate a new site-to-site channel. Later, once QKD is operational and a secure quantum channel exists, a pre-shared key is constructed from quantum key material to replace the public cryptographic key.
\item Post-quantum cryptography, in the form of public-key algorithms that are considered secure against attack by quantum computers, may be utilized to boot-strap authentication in a fashion similar to the previous option; a public-key infrastructure or a system such as Kerberos may be required.
\end{enumerate}

Sites must also communicate at the link layer over a classical channel to carry out the discussion required in the QKD protocol. The above approaches may be used to ensure unconditionally secure authentication of this channel. Note that these options only apply to the initial authentication, and encryption if appropriate, of a new communications channel between sites at the appropriate layer; they do not apply to host communication, which relies on QKD operation and the issuance of QKD-generated key material through a key establishment protocol as described in section~\ref{sec:kms_layer}, in order to conduct symmetric key encryption of user data. 

\subsection{Security of the System}

The design of the KMS assumes the satisfaction of certain security properties, including the following:

\begin{enumerate}
\item The encoding and decoding devices within the network layer are physically secure from attack.
\item The conventional channel used for coordination of the QKD protocol between parties may be read by an eavesdropper. However, an attacker cannot modify the contents of it or inject unauthorized contents into it without being detected. Thus, the conventional network channel must be authenticated, and there is a feasible method of doing so.
\item Once a quantum computer capable of mounting a successful attack against the security of a protocol such as TLS exists, any random number generator that is employed must be truly random. A Quantum Random Number Generator (QRNG) may be used for the purpose of generating a random nonce to prevent a replay attack during key negotiation in such protocols as TLS. Otherwise, a pseudo-random number generator may suffice.
\end{enumerate}

The proposed host-based protocol described in Section~\ref{subsec:key_issuance} provides confidentiality of user data by encrypting it using the symmetric session key, once it is successfully negotiated and secure communication has begun between hosts of separate sites. The only relevant traffic transmitted over the conventional network prior to the secure communication is the remote session key negotiation message sent from the originator to the recipient, and the remote session confirmation reply-back. An attacker Eve will be unable to mount a successful man-in-the-middle attack. If Eve intercepts the remote session key negotiation message from Alice to Bob, then the encrypted key selection packet cannot be read, as it is encrypted with an inter-site key. Eve cannot successfully replay this message to Bob, since authenticated encryption is utilized. The key negotiation message should be protected by an authentication tag such as a Message Authentication Code (MAC) as part of a method such as encrypt-then-MAC. This method ensures that Bob will only read valid messages from Alice, and it makes them unforgeable under chosen-ciphertext attack (where Eve chooses to decrypt known ciphertexts and recover the key from the resulting plaintexts). The MAC can be applied using additional key material from the quantum key pool, which may be retrieved at the same time as the key material intended for the host-to-host session key itself. Encrypt-then-MAC has been prescribed for TLS \cite{rfc7366}. 

The session key is changed at regular intervals with the assumption that it will increase security. If a session key is successfully defeated, then only the communication encrypted with that session key will be unlocked; the protocol thus achieves perfect forward secrecy, as the session keys that are assigned are independent of each other. The compromise of one session key will compromise only the messages encrypted with that session key, and not messages belonging to other sessions that were encrypted with different session keys.

The QKD technology prevents key material from being made accessible by the attacker. Suppose that an attacker Eve attempts to interfere with the protocol. Eve will be unable to access the quantum key pools, as they are externally inaccessible, and will thus be unable to read the QKD-generated key material. Furthermore, Eve will be unable to contribute any invalid key material to the pools, as the QKD process cannot be successfully interfered with. Also, the KMS is accessible only within its own site through its published external interface API; it is not externally accessible from parties outside of its network domain and firewall. The KMS operates within a physically secured environment that protects physical access to the computer hardware responsible for the KMS and quantum key pool storage.

The policy engine enforces rules for the issuance and use of key material. The policy engine can enforce what cipher to use for host communication, the minimum key length, and the usable lifetime of the session key.  Furthermore, the policy engine can validate a remote host before issuing the session key. Different hosts on different sites may be associated with different sets of permissions. These permissions may indicate the required levels of security for host-to-host communication to be carried out.

When utilizing the session keys issued by the KMS to encrypt host traffic, a suitable symmetric-key algorithm can be utilized, such as the Advanced Encryption Standard (AES) \cite{fips197}. The cipher will be initialized with the session key issued by the KMS. The cipher will provide privacy of host information. The security schemes presented do not rely upon public-key cryptography, which is considered vulnerable to successful attack by a quantum computer. The schemes rely upon pre-shared symmetric keys only; the keys are generated on both hosts through QKD, which is considered secure based on the laws of quantum physics.

\subsection{Combination with Post-Quantum Algorithms}
\label{sec:combination_post_quantum}

It is important to note that exclusive reliance on QKD may not be justifiable. In fact, QKD augments well the conventional post-quantum ecosystem; the aim is not to have QKD replace it. QKD protocols are fundamentally and provably secure against cryptanalysis and keys remain secure forever from the time of creation, as the technology is based on the laws of quantum physics; however, practical implementations of QKD may possess vulnerabilities. The principal problem in practice is that QKD requires special hardware and has limited key rate over long distance, so that it is likely not suitable for providing key material for one-time pad encryption for large volumes of data in an enterprise setting. Post-quantum public key cryptography\footnote{To avoid ambiguity, the term ``quantum-safe algorithms" is not used, as QKD is also considered to be quantum-safe.}, on the other hand, requires no new infrastructure, and works within a Public Key Infrastructure (PKI). Both QKD protocols and post-quantum algorithms are an active research area and there is a long road ahead to build confidence in primitives and implementations, and to define standards; thus, they are not yet ready for widespread commercial use.

A possible key construction strategy to diversify risk is to combine the two techniques:

\begin{enumerate}
\item Firstly, as suggested in section~\ref{sec:authentication_sites}, quantum-safe algorithms authenticate the QKD channel via PKI. In this case, future cryptanalysis of the post-quantum authentication does not compromise the security of the QKD key established before the post-quantum algorithm was broken. In other words, one only relies on the short-term security of the post-quantum algorithm.
\item Secondly, post-quantum algorithms and QKD are used independently to create two session keys; the two keys are combined, such as through an exclusive~OR operation, so that an attacker must break both. In this case, if either the QKD or the post-quantum key generation become vulnerable, the overall security is not compromised. Furthermore, each type of key generation would certainly have to fail in a different way. This solution is consistent with the National Institute of Standards and Technology (NIST) approving the use of quantum-resistant cryptography in FIPS-140 approved systems, as long as the hybrid solution includes one FIPS-validated algorithm {\cite{nist17}}.
\end{enumerate}

\section{Lessons Learned and Recommendations}
\label{sec:recommendations}

We propose that our four-layered architecture is a valuable framework for the QKD community to facilitate the advancement and adoption of QKD in practical network systems. There is a myriad of QKD technologies that are still evolving; a flexible framework can accommodate any QKD technology without rework, and disruption to the entire tool chain is avoided. End-users may design their applications to obtain QKD-generated keys from a KMS without adapting to changes in the inner workings of the KMS, or the QNL algorithms and topology, or the underlying QLL technologies. 

Our design is not generic in nature; rather, it is highly customized for the challenges of QKD. It maximizes the security benefits of QKD while mitigating its limitations. Because the quantum key generation rate is still relatively lower than in conventional cryptosystems, there needs to be flexibility in how quantum key material is consumed; the policy engine can dictate various key consumption strategies. Since hosts may wish to communicate across the entire network, it is useful to have intermediate nodes assist with key generation. Because the demand for keys may continually change, the system must adapt by deciding when to run QKD, using what paths in the network, and for how long. 

The size of investment in QKD technology and infrastructure is an important consideration. To minimize the switching cost and to speed up adoption, it may be best to integrate with existing security and authentication protocols and standards. The use of shared infrastructure, such as multiplexing the QKD protocol with conventional user traffic on a shared fibre-optic line, is another tenet of rationalization and fast ramp-up.

Our recommendations for designing and implementing a QKD-based system in the future would be to design in a manner that is agnostic to the underlying QKD technology. Running the system as a set of micro-services will provide plug-and-play modularity within the layered architecture. The use of both QKD and quantum-safe algorithms enables sufficient session keys to be generated for a large host population that is governed by policy rules.

Finally, it is important to create a system that can effectively scale and support popular use cases. Optimizing the key generation activity based on real-time demand monitoring ensures maximum utilization of the security infrastructure. Defining flexible security levels and associated key issuance strategies will result in more efficient consumption of keys. 

A useful performance metric is the key generation rate for an arbitrary pair of nodes that are not directly connected. The rate is upper-bounded by the implemented QKD technology, but the network designer has the opportunity to improve performance to a degree through sound network engineering. Possibilities include reducing the number of hops in the paths between source and destination pairs experiencing higher demand, increasing the capacity of key quantum links, and utilizing more optimal network topologies that are highly meshed. Agile networking such as software-defined networks with centrally-managed intelligence and control can help optimize in this regard.

\ack{\small This work was supported by Quantum Canada and was performed in collaboration with the National Research Council of Canada.}

\section*{References}

\hbadness 10000 % Remove underfull \hbox warnings in bilbiography
\bibliographystyle{IEEEtran}
\bibliography{qkd}

\begin{thebibliography}{10}
\providecommand{\url}[1]{#1}
\csname url@samestyle\endcsname
\providecommand{\newblock}{\relax}
\providecommand{\bibinfo}[2]{#2}
\providecommand{\BIBentrySTDinterwordspacing}{\spaceskip=0pt\relax}
\providecommand{\BIBentryALTinterwordstretchfactor}{4}
\providecommand{\BIBentryALTinterwordspacing}{\spaceskip=\fontdimen2\font plus
\BIBentryALTinterwordstretchfactor\fontdimen3\font minus
  \fontdimen4\font\relax}
\providecommand{\BIBforeignlanguage}[2]{{%
\expandafter\ifx\csname l@#1\endcsname\relax
\typeout{** WARNING: IEEEtran.bst: No hyphenation pattern has been}%
\typeout{** loaded for the language `#1'. Using the pattern for}%
\typeout{** the default language instead.}%
\else
\language=\csname l@#1\endcsname
\fi
#2}}
\providecommand{\BIBdecl}{\relax}
\BIBdecl

\bibitem{diamanti16}
E.~Diamanti, H.-K. Lo, B.~Qi, and Z.~Yuan, ``Practical challenges in quantum
  key distribution,'' \emph{Quantum Information}, November 2016.

\bibitem{sasaki17}
M.~Sasaki, ``Quantum networks: where should we be heading?'' \emph{Quantum
  Science and Technology}, vol.~2, no.~2, 2017.

\bibitem{mink09}
A.~Mink, S.~Frankel, and R.~Perlner, ``{Quantum Key Distribution (QKD) and
  Commodity Security Protocols: Introduction and Integration},''
  \emph{International Journal of Network Security \& Its Applications (IJNSA)},
  vol.~1, no.~2, July 2009.

\bibitem{elboukhari10}
M.~Elboukhari, M.~Azizi, and A.~Azizi, ``{Improving TLS Security By Quantum
  Cryptography},'' \emph{International Journal of Network Security \& Its
  Applications (IJNSA)}, vol.~2, no.~3, 2010.

\bibitem{assche06}
G.~V. Assche, \emph{{Quantum Cryptography and Secret-Key Distillation}}.\hskip
  1em plus 0.5em minus 0.4em\relax Cambridge University Press, 2006.

\bibitem{pattaranantakul12}
M.~Pattaranantakul, A.~Janthong, K.~Sanguannam, P.~Sangwongngam, and
  K.~Sripimanwat, ``{Secure and Efficient Key Management Technique in Quantum
  Cryptography Network},'' in \emph{Fourth International Conference on
  Ubiquitous and Future Networks (ICUFN)}, July 2012.

\bibitem{pattaranantakul15}
M.~Pattaranantakul, K.~Sanguannam, P.~Sangwongngam, and C.~Vorakulpipat,
  ``{Efficient Key Management Protocol for Secure RTMP Video Streaming toward
  Trusted Quantum Network},'' \emph{Electronics and Telecommunications Research
  Institute (ETRI) Journal}, vol.~37, no.~4, August 2015.

\bibitem{kmip09}
\emph{{Key Management Interoperability Protocol (KMIP): Addressing the Need for
  Standardization in Enterprise Key Management}}, {Organization for the
  Advancement of Structured Information Standards (OASIS)}, May 2009.

\bibitem{murali15}
G.~Murali and R.~S. Prasad, ``{Quantum Cryptography Based Solution for Secure
  and Efficient Key Management for E-Governance in India},''
  \emph{International Conference on Applied and Theoretical Computing and
  Communication Technology (iCATccT)}, pp. 18--26, October 2015.

\bibitem{bristol17}
\BIBentryALTinterwordspacing
{University of Bristol}. {Quantum in the Cloud}. [Online]. Available:
  \url{{http://www.bristol.ac.uk/physics/research/quantum/engagement/qcloud/}}
\BIBentrySTDinterwordspacing

\bibitem{peev09}
M.~Peev, C.~Pacher, R.~All{\'e}aume, C.~Barreiro, J.~Bouda, W.~Boxleitner,
  T.~Debuisschert, E.~Diamanti, M.~Dianati, J.~F. Dynes, S.~Fasel, S.~Fossier,
  M.~F{\"u}rst, J.-D. Gautier, O.~Gay, N.~Gisin, P.~Grangier, A.~Happe,
  Y.~Hasani, M.~Hentschel, H.~H{\"u}bel, G.~Humer, T.~L{\"a}nger, M.~Legr{\'e},
  R.~Lieger, J.~Lodewyck, T.~Lor{\"u}nser, N.~L{\"u}tkenhaus, A.~Marhold,
  T.~Matyus, O.~Maurhart, L.~Monat, S.~Nauerth, J.-B. Page, A.~Poppe,
  E.~Querasser, G.~Ribordy, S.~Robyr, L.~Salvail, A.~W. Sharpe, A.~J. Shields,
  D.~Stucki, M.~Suda, C.~Tamas, T.~Themel, R.~T. Thew, Y.~Thoma, A.~Treiber,
  P.~Trinkler, R.~Tualle-Brouri, F.~Vannel, N.~Walenta, H.~Weier,
  H.~Weinfurter, I.~Wimberger, Z.~L. Yuan, H.~Zbinden, and A.~Zeilinger, ``{The
  SECOQC quantum key distribution network in Vienna},'' \emph{New Journal of
  Physics}, vol.~11, July 2009.

\bibitem{tajima17}
A.~Tajima, T.~Kondoh, T.~Ochi, M.~Fujiwara, K.~Yoshino, H.~Iizuka, T.~Sakamoto,
  A.~Tomita, E.~Shimamura, and S.~Asami, ``Quantum key distribution network for
  multiple applications,'' \emph{Quantum Science and Technology}, 2017.

\bibitem{hughes13}
R.~J. Hughes, J.~E. Nordholt, K.~P. McCabe, R.~T. Newell, C.~G. Peterson, and
  R.~D. Somma, ``{Network-Centric Quantum Communications with Application to
  Critical Infrastructure Protection},'' \emph{{Quantum Physics}}, May 2013.

\bibitem{chapuran09}
T.~E. Chapuran, P.~Toliver, N.~A. Peters, J.~Jackel, M.~S. Goodman, R.~J.
  Runser, S.~R. McNown, N.~Dallmann, R.~J. Hughes, K.~P. McCabe, J.~E.
  Nordholt, C.~G. Peterson, K.~T. Tyagi, L.~Mercer, and H.~Dardy, ``Optical
  networking for quantum key distribution and quantum communications,''
  \emph{New Journal of Physics}, vol.~11, October 2009.

\bibitem{alshamsi05}
A.~Alshamsi and T.~Saito, ``{A technical comparison of IPSec and SSL},''
  \emph{19th IEEE International Conference on Advanced Information Networking
  and Applications (AINA)}, March 2005.

\bibitem{etsi2015}
``{Quantum Safe Cryptography and Security},'' ETSI (European Telecommunications
  Standards Institute), White Paper~8, June 2015.

\bibitem{fips197}
\emph{{Federal Information Processing Standards Publication 197: Announcing the
  Advanced Encryption Standard (AES)}}, National Institute of Standards and
  Technology, November 2001.

\bibitem{salvail10}
L.~Salvail, M.~Peev, E.~Diamanti, R.~All{\'e}aume, N.~L{\"u}tkenhaus, and
  T.~L{\"a}nger, ``Security of trusted repeater quantum key distribution
  networks,'' \emph{Journal of Computer Security}, vol.~18, no.~1, January
  2010.

\bibitem{shamir79}
A.~Shamir, ``How to share a secret,'' \emph{Communications of the ACM},
  vol.~22, no.~11, November 1979.

\bibitem{etsi2010}
``{ETSI GS QKD 004 V1.1.1 Quantum Key Distribution (QKD) Application
  Interface},'' ETSI, Tech. Rep., December 2010.

\bibitem{padamvathi16}
V.~Padamvathi, B.~V. Vardhan, and A.~V.~N. Krishna, ``{Quantum Cryptography and
  Quantum Key Distribution Protocols: A Survey},'' in \emph{IEEE 6th
  International Conference on Advanced Computing (IACC)}, February 2016.

\bibitem{lo15}
H.-K. Lo, M.~Curty, and K.~Tamaki, ``Secure quantum key distribution,''
  \emph{Nature Photonics}, vol.~8, no.~8, 8 2014.

\bibitem{pirandola16}
S.~Pirandola and S.~L. Braunstein, ``{Physics: Unite to build a quantum
  Internet},'' \emph{Nature}, vol. 532, no. 7598, April 2016.

\bibitem{rfc4279}
\BIBentryALTinterwordspacing
``{Pre-Shared Key Ciphersuites for Transport Layer Security (TLS), RFC 4279.}''
  December 2005. [Online]. Available: \url{https://tools.ietf.org/html/rfc4279}
\BIBentrySTDinterwordspacing

\bibitem{rfc5246}
\BIBentryALTinterwordspacing
``{The Transport Layer Security (TLS) Protocol Version 1.2, RFC 5426.}'' August
  2008. [Online]. Available: \url{https://tools.ietf.org/html/rfc5246}
\BIBentrySTDinterwordspacing

\bibitem{rfc2409}
\BIBentryALTinterwordspacing
``{The Internet Key Exchange (IKE), RFC 2409.}'' November 1998. [Online].
  Available: \url{https://tools.ietf.org/html/rfc2409}
\BIBentrySTDinterwordspacing

\bibitem{rfc4120}
\BIBentryALTinterwordspacing
``{Kerberos Network Authentication Service (V5), RFC 4120.}'' July 2005.
  [Online]. Available: \url{https://www.ietf.org/rfc/rfc4120.txt}
\BIBentrySTDinterwordspacing

\bibitem{ferguson10}
N.~Ferguson, B.~Schneier, and T.~Kohno, \emph{{Cryptography
  Engineering}}.\hskip 1em plus 0.5em minus 0.4em\relax Wiley Publishing, 2010.

\bibitem{tanenbaum2011}
A.~S. Tanenbaum and D.~J. Wetherall, \emph{Computer Networks}, 5th~ed.\hskip
  1em plus 0.5em minus 0.4em\relax Prentice Hall, 2011.

\bibitem{rfc2328}
\BIBentryALTinterwordspacing
``{OSPF (Open Shortest Path First) Version 2, RFC 2328.}'' April 1998.
  [Online]. Available: \url{https://tools.ietf.org/html/rfc2328}
\BIBentrySTDinterwordspacing

\bibitem{kurose13}
J.~F. Kurose and K.~W. Ross, \emph{{Computer Networking: A Top-Down Approach,
  6th Edition}}.\hskip 1em plus 0.5em minus 0.4em\relax Pearson Education,
  2013.

\bibitem{girard90}
A.~Girard, \emph{{Routing and Dimensioning in Circuit-Switched
  Networks}}.\hskip 1em plus 0.5em minus 0.4em\relax Addison-Wesley Longman
  Publishing, 1990.

\bibitem{karakostas08}
G.~Karakostas, ``{Faster approximation schemes for fractional multicommodity
  flow problems},'' \emph{ACM Transactions on Algorithms (TALG)}, vol.~4,
  no.~1, March 2008.

\bibitem{fleischer00}
L.~K. Fleischer, ``{Approximating Fractional Multicommodity Flow Independent of
  the Number of Commodities},'' \emph{SIAM Journal on Discrete Mathematics},
  vol.~13, no.~4, 2000.

\bibitem{shreedhar95}
M.~Shreedhar and G.~Varghese, ``{Efficient fair queueing using deficit round
  robin},'' in \emph{{SIGCOMM '95: Proceedings of the conference on
  applications, technologies, architectures, and protocols for computer
  communication}}, 1995.

\bibitem{salvail09}
L.~Salvail, M.~Peev, E.~Diamanti, R.~All\'eaume, N.~L\"utkenhaus, and
  T.~L\"anger, ``{Security of Trusted Repeater Quantum Key Distribution
  Networks},'' \emph{Journal of Computer Security}, vol.~18, no.~1, January
  2010.

\bibitem{rfc7366}
\BIBentryALTinterwordspacing
``{Encrypt-then-MAC for Transport Layer Security (TLS) and Datagram Transport
  Layer Security (DTLS), RFC 7366.}'' September 2014. [Online]. Available:
  \url{https://tools.ietf.org/html/rfc7366}
\BIBentrySTDinterwordspacing

\bibitem{nist17}
\BIBentryALTinterwordspacing
``{NIST Cryptographic Technology Group: Post-Quantym Crypto Standardization
  FAQ},'' April 2017. [Online]. Available:
  \url{http://csrc.nist.gov/groups/ST/post-quantum-crypto/faq.html}
\BIBentrySTDinterwordspacing

\end{thebibliography}

\end{document}